\documentclass[twocolumn,notitlepage,showpacs,prl,preprintnumbers,amsmath,amssymb,aps,floatfix]{revtex4-1}
\usepackage{graphicx}
\usepackage{dcolumn}
\usepackage{bm}
\usepackage{soul}
\usepackage{color}
\usepackage{lineno,hyperref}
\usepackage{graphicx}
\usepackage{dcolumn}
\usepackage{bm}
\usepackage{enumerate}
\usepackage[dvips]{epsfig}
\usepackage[ center ]{ subfigure }
\usepackage{xcolor}
\usepackage{amsmath}
\usepackage{soul}
\usepackage{color}

\newcommand{\BEQ}{\begin{equation}}
\newcommand{\EEQ}{\end{equation}}
\newcommand{\BEA}{\begin{eqnarray}}
\newcommand{\EEA}{\end{eqnarray}}

\newcommand{\nn}{\nonumber}

\begin{document}
\title{ Noise-induced resonant acceleration of a  charge  in an intermittent magnetic field: an exact solution for ergodic and non-ergodic fluctuations }

\author{Gerardo Aquino$^{1}$}
\email{Gerardo.Aquino@apha.gov.uk}
\author{Mauro Bologna$^{2}$}
\affiliation{$^{1}$Department of Computing,  Goldsmiths, University of London, London, UK}
\affiliation{$^{2}$Departamento de Ingenier\'ia El\'ectrica-Electr\'onica, Universidad de Tarapac\'a,  Arica 1000000}


\begin{abstract}
We study the diffusion of a charged particle in a magnetic field subject to stochastic dichotomous fluctuations.
The associated induced electric field gives rise to non-trivial dynamical regimes.
In particular, when the mean magnetic field vanishes, the particle remains confined within a finite radius, regardless of the fluctuation statistics.
For a non-zero mean field, we show—using a density approach for Poissonian fluctuations—that the particle undergoes an exponential regime of accelerated diffusion.
Crucially and more generally, adopting a trajectory-based formalism, we derive an exact analytical solution valid for arbitrary waiting-time distributions, including non-Poissonian and non-ergodic cases.
Even rare, abrupt field reversals are shown to trigger exponential acceleration of the particle’s diffusion.
We demonstrate that this behaviour stems from noise exciting resonance bands present for periodic fluctuations, and we propose noise-induced resonant acceleration as a robust and efficient charge acceleration mechanism, potentially more effective than Fermi’s classic model for cosmic-ray acceleration.
 \end{abstract}

\maketitle

The origin of high-energy particles in the universe remains an open question.
Observational evidence indicates that they are largely produced by explosive astrophysical phenomena.
Since Fermi’s pioneering work \cite{Fermi} and later work on their possible origin, energy spectrum  and wave-particle interaction \cite{arbell, davids,trub}, several theoretical explanations \cite{drury, vulpe,moine} and experimental validations have been attempted \cite{an, tang, comm},  yet,  the precise acceleration mechanisms are still debated. Importantly  intermittency of the magnetic field has been mostly overlooked \cite{interm, interm2}   despite accruing experimental evidence. 
Motivated by these findings, and by the known influence of magnetic fluctuations on particle diffusion  ~\cite{marc,a} and plasma confinement~\cite{shun},
we present a comprehensive solution to the problem of a charged particle diffusing in an intermittent magnetic field.
We investigate how such fluctuations can energise particles, providing a mechanism that, without accounting for collective effects, applies to acceleration at low densities and is relevant to both space and laboratory plasmas, and in all cases where  intermittent magnetic fields occur naturally or are imposed artificially.

Within a diffusive framework \cite{met,shun} and overcoming the limitations of previous work \cite{aq3}, we offer a full analytical treatment that explicitly incorporates the induced electric field.
We find two main dynamical regimes:
(i) confinement when the average magnetic field vanishes, and
(ii) exponentially accelerated diffusion when a finite mean field is present.

The acceleration arises from a resonance between magnetic-field fluctuations and the particle’s orbital frequency, and it persists even under rare, abrupt field changes. 

The article  is organised as follows: 
 we first define the stochastic equation driving the diffusive process;   we then study it analytically,  first in the case of Poissonian fluctuations within a density approach,   and then, within a trajectory-based approach, we obtain a complete exact solution for  a generic waiting times distribution of the fluctuations. The two approaches produce the same solution in the Poissonian case and are also confirmed by numerical simulations.
  We then provide an explanation of the dynamics based
 on the property of the case of periodic fluctuations of   ${\bf B}$.

We start from the equation of motion for a charged particle moving in a region with electric and magnetic fields~\cite{land8,jak}:
\begin{eqnarray} \label{f0}
m \dot{\mathbf{v}} = q \left( \mathbf{v} \times \mathbf{B} + \mathbf{E} \right),
\end{eqnarray}
where $m$ is the particle mass, $q$ its charge, $\mathbf{v}$ the velocity, and $\dot{\mathbf{v}} \equiv d\mathbf{v}/dt$.  
The magnetic field  $\mathbf{B}$  is taken as a uniform background $B_0$ along the $z$-axis plus a randomly fluctuating component $B_1$ along the same axis:
\[
\mathbf{B}(t) = B(t)\,\mathbf{k} = [\,B_0 + B_1 \xi(t)\,]\mathbf{k},
\]
where $\xi(t)$ is a stochastic dichotomous variable taking the values $\pm1$.  
The sojourn time in each state is a random variable with distribution $\psi(t)$.  
We consider two cases:  
(i) an exponential law, $\psi(t) = \gamma e^{-\gamma t}$ (Poisson statistics), and  
(ii) a power law, $\psi(t) \propto t^{-\alpha-1}$ with $0 < \alpha < 2$ (non-Poissonian statistics).

According to Faraday–Lenz’s law, the induced electric field is approximately
\begin{equation}\label{electr1}
\mathbf{E}(t) = \frac{\dot{B}(t)}{2}\left( y\mathbf{i} - x\mathbf{j} \right).
\end{equation}
This approximation is well-known in the literature (see Ref. \cite{kavorkia} as a textbook example). Furthermore, by assuming the uniform magnetic field component generated by a solenoid and using the exact approach of Ref. \cite{bolo} for an arbitrary flowing current density, it is possible to derive an expression for the electric and magnetic fields near the solenoid axis (i.e., the z-axis of our reference system) that coincides with the paper's approximation. 

Inserting Eq.~\eqref{electr1} into Eq.~\eqref{f0} yields the coupled equations:
\begin{eqnarray} \label{f1_e}
&&\ddot{x}=  \omega(t) \dot{y}+ \dot{\omega}(t) y/2
\\\label{f2_e}
&&\ddot{y}=- \omega(t)\dot{x}- \dot{\omega}(t) x/2
\end{eqnarray}

where the instantaneous Larmor frequency is
\[
\omega(t) = \frac{q B(t)}{m} = \omega_0 + \omega_1 \xi(t), \qquad 
\text{with} \;  \; \; \omega_{0,1} = \frac{q B_{0,1}}{m}.
\]

Introducing the complex variable $z(t) = x(t) + i y(t)$, Eqs.~\eqref{f1_e}–\eqref{f2_e} reduce to a single stochastic differential equation:
\begin{equation}\label{electr2}
\ddot{z} = -i \omega(t) \dot{z} - \frac{i}{2} \dot{\omega}(t) z.
\end{equation}
Following the approach of Refs.~\cite{gitt2,gitt}, we perform the transformation
\begin{equation}\label{electr3}
z(t) = \exp\!\left[-\frac{i}{2} \int^t \omega(s)\,ds\right] u(t),
\end{equation}
which leads to
\begin{equation}\label{electr4}
\ddot{u}(t) + \frac{\omega^2(t)}{4}\,u(t) = 0.
\end{equation}
In the next sections we analyse Eq.~\eqref{electr4}, and therefore Eq.~\eqref{electr2}, for different statistics of the fluctuating field.

\subsection{Numerical analysis}
To simulate the particle dynamics, we generate a sequence of random time intervals drawn from the chosen distribution $\psi(t)$.  
Within each interval, the magnetic field remains constant, and the particle performs uniform circular motion determined by the current field value.  
At each instantaneous switch of the field, the change in the particle velocity due to the induced electric field is obtained by integrating Eqs.~\eqref{f1_e} and~\eqref{f2_e} over a small interval $\Delta t$.  
Taking the limit $\Delta t \to 0$ gives
\begin{eqnarray}\label{kicksa}
\Delta v_x &=& \int_t^{t+\Delta t} \!\! dt'
\left[ \omega(t') \dot{y}(t') + \frac{\dot{\omega}(t')}{2} y(t') \right]
   = \pm \omega_1\, y, \\[1ex]
\label{kicksb}
\Delta v_y &=& \int_t^{t+\Delta t} \!\! dt'
\left[ -\omega(t') \dot{x}(t') - \frac{\dot{\omega}(t')}{2} x(t') \right]
   = \mp \omega_1\, x,
\end{eqnarray}
since $\dot{\omega}(t) = \pm 2\omega_1 \delta(t)$.  
After each switch, the centre and radius of the subsequent circular trajectory are recomputed using the updated velocity.

Representative trajectories obtained with this algorithm are shown in Fig.~1.  
To evaluate diffusion properties, such as the mean square displacement $\langle r^2(t) \rangle$, we average over many independent realisations of these trajectories.
{
 \begin{figure}[h]
 \centering
 \includegraphics[scale=.15]{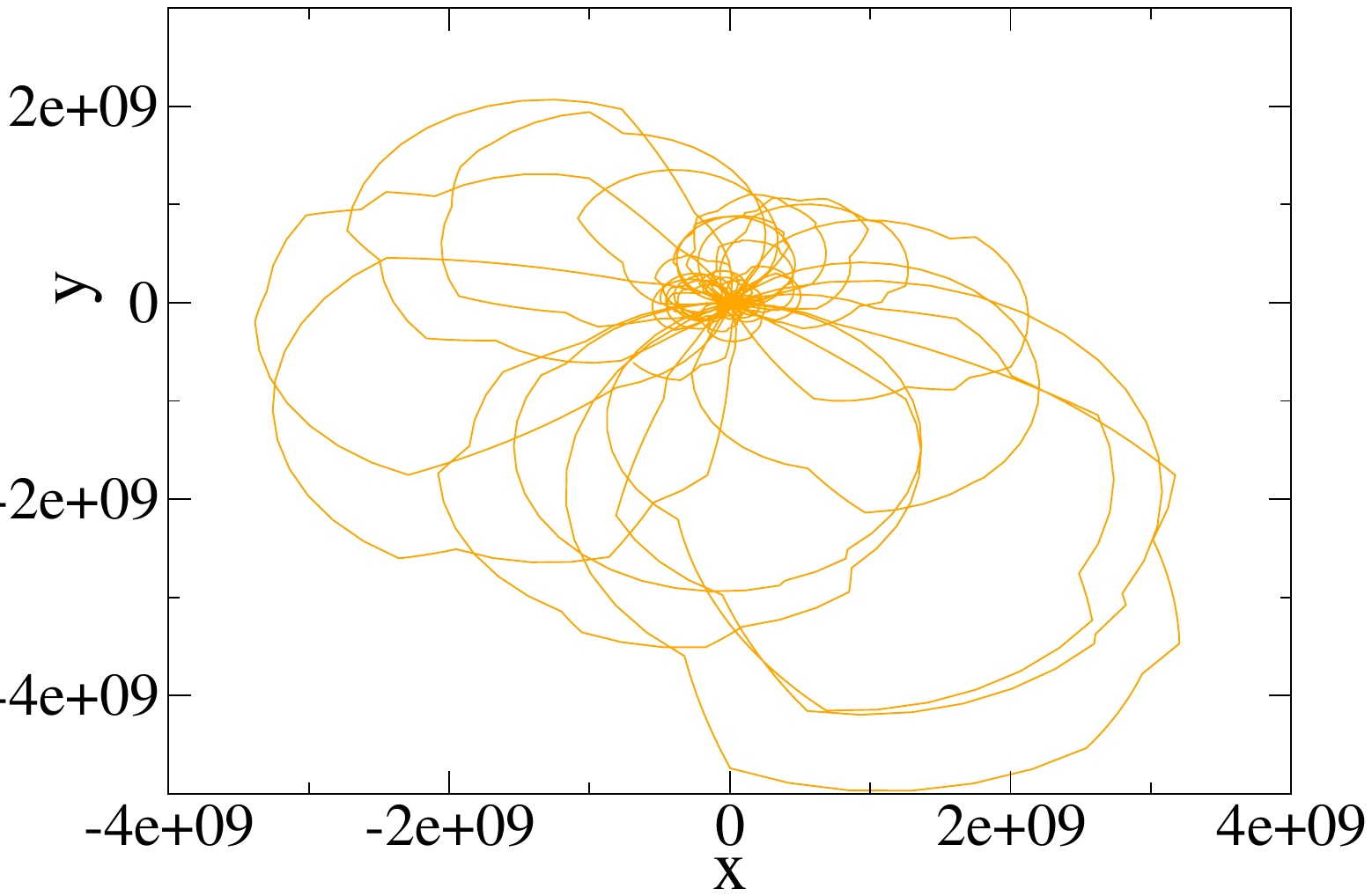} 
   \includegraphics[scale=.15]{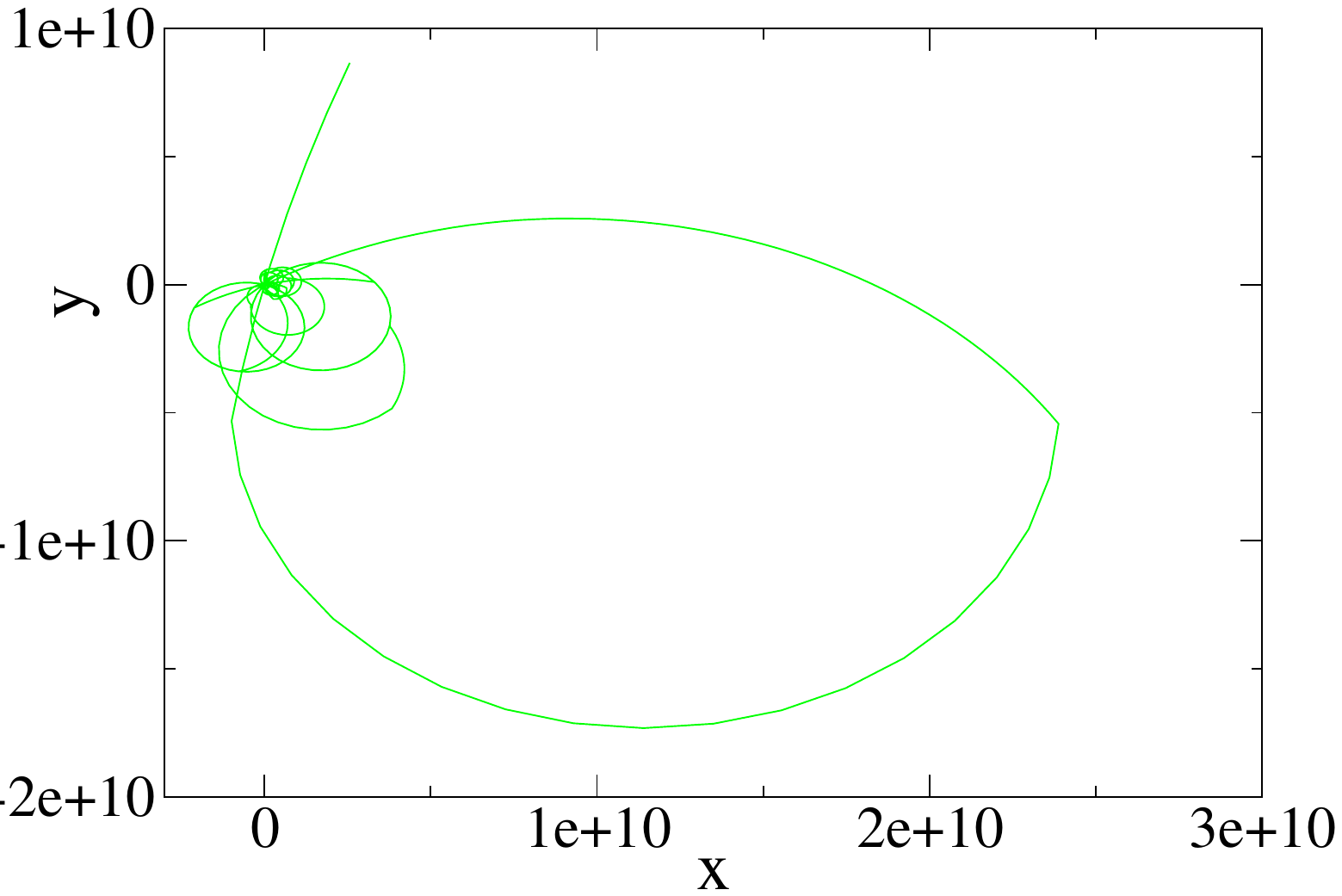} \\
         
 \includegraphics[scale=0.15]{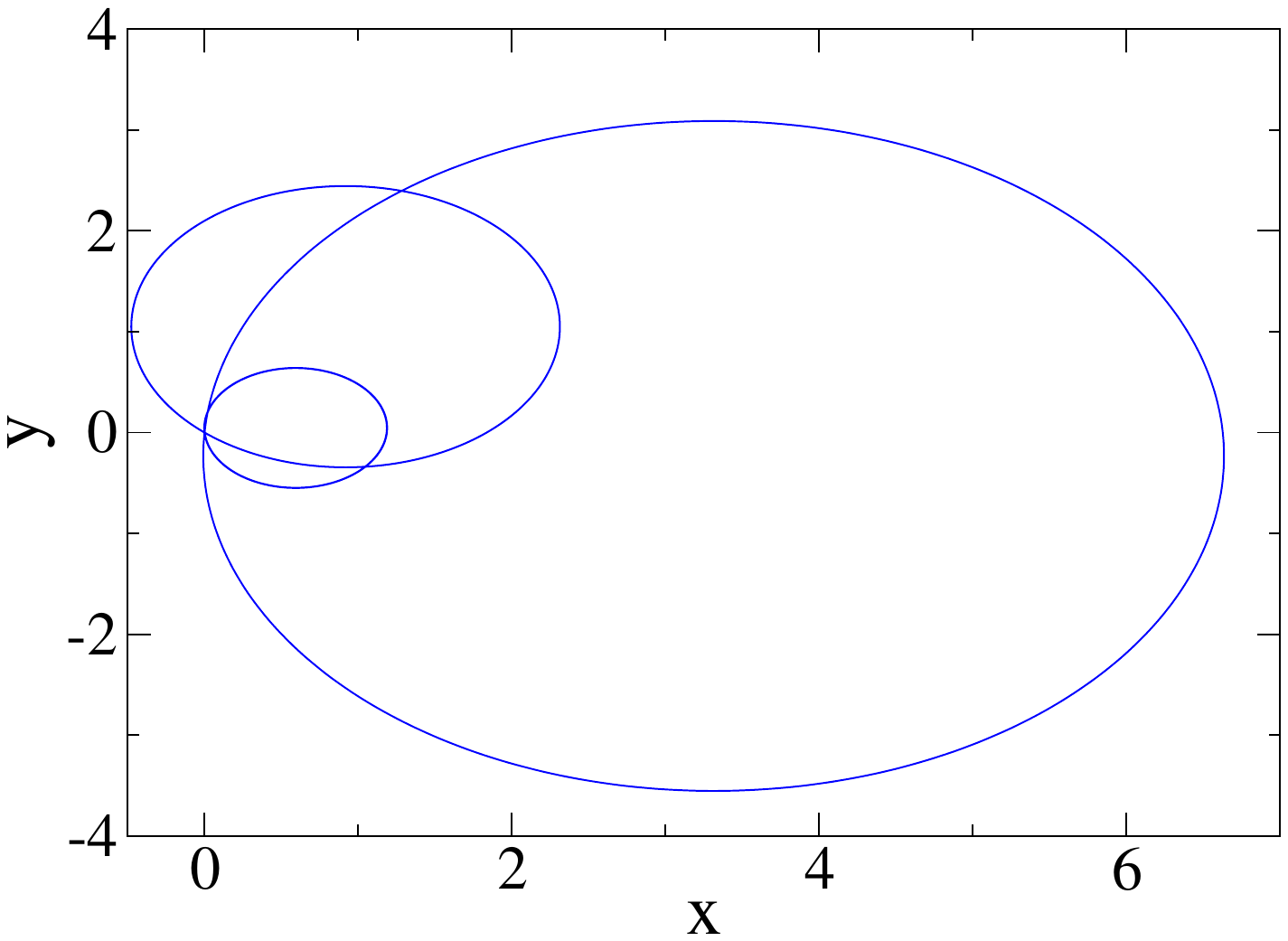} 
              \includegraphics[scale=.15]{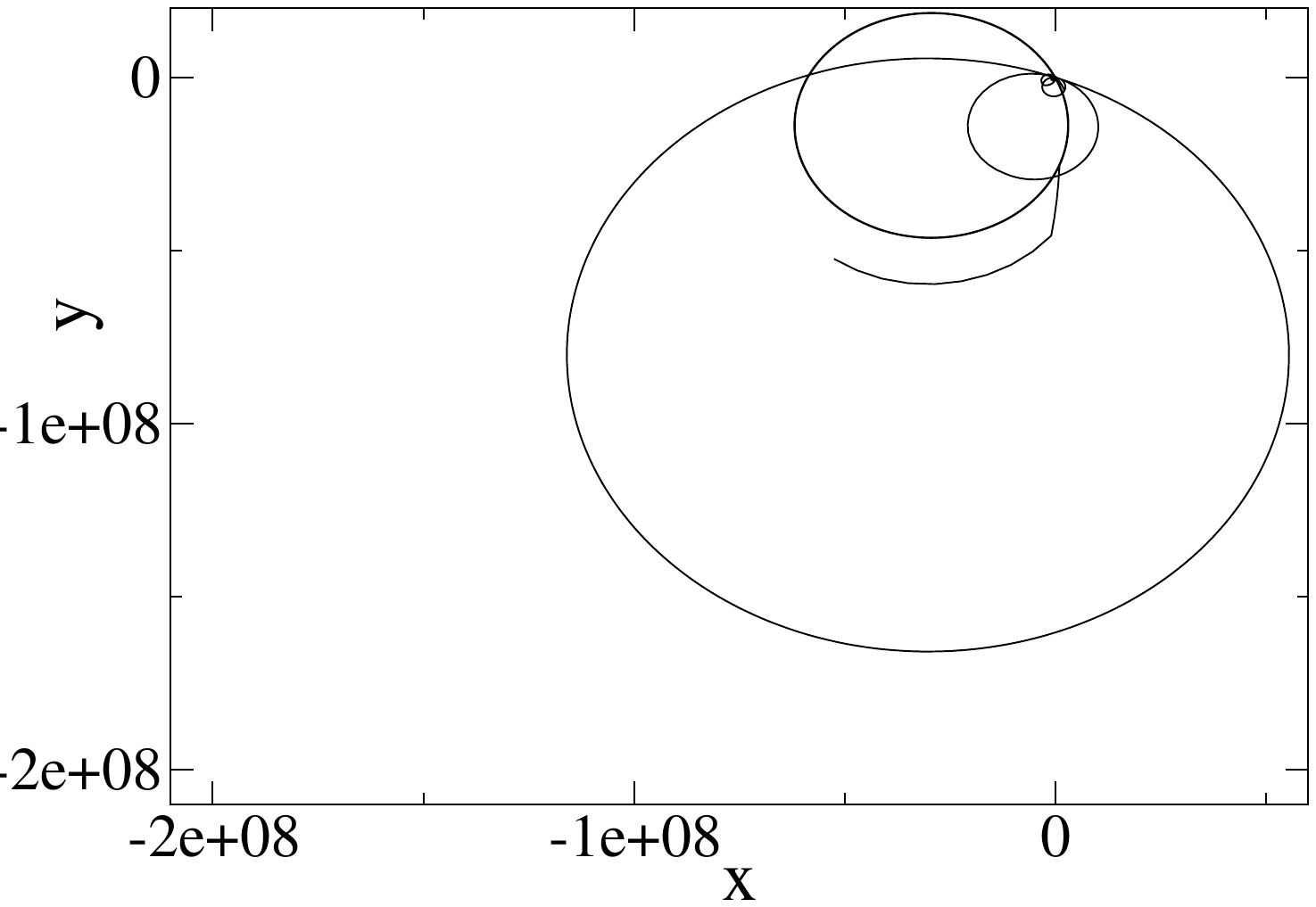} 
    \caption{Examples of trajectories generated with $\omega_0=1$ and $\omega_1=0.7$ at $t=10000$dt with dt$=0.1$. Top runs are generated with Poisson distribution $\psi(t)=\gamma e^{-\gamma t}$ with  $\gamma=0.1$ (left)  and $\gamma=0.5$  (right). Bottom runs  with power-law distribution $\psi(t )\sim t^{-\alpha-1}$ with $\alpha=0.25$ (left) and $\alpha=0.75$ (right)}
    \label{fig:subfigures}
  \end{figure}
 } 

\section{Density approach}

For Poissonian fluctuations, the full dynamics of the process can be solved using a density-based method.  
In particular, the Shapiro–Loginov differentiation formula~\cite{log} is applicable.  
For a dichotomous process $\xi(t)$ whose $n$-point correlation function satisfies
\[
\frac{\partial}{\partial t}
   \langle \xi(t)\,\xi(t_1)\cdots\xi(t_n) \rangle
   = -\gamma
   \langle \xi(t)\,\xi(t_1)\cdots\xi(t_n) \rangle,
\]
the following differentiation rule holds:
\begin{equation}\label{SLformula}
\frac{\partial}{\partial t}
   \langle \xi(t)\,\rho(t) \rangle
   = -\gamma \langle \xi(t)\,\rho(t) \rangle
     + \Big\langle \xi(t)\, \frac{\partial \rho(t)}{\partial t} \Big\rangle,
\end{equation}
where the average is taken over all realisations of $\xi(t)$.  

Equation~\eqref{SLformula} can be employed, in the Poissonian case, to obtain the mean value $\langle z(t) \rangle$, and consequently its real and imaginary parts, $\langle x(t) \rangle$ and $\langle y(t) \rangle$.  
Of particular interest is the mean square displacement
\[
\langle r^2(t) \rangle = \langle x^2(t) + y^2(t) \rangle,
\]
which provides a direct measure of the diffusion process.  
The detailed derivation of these results is presented in Appendix~A3, and the corresponding behaviour of $\langle r^2(t) \rangle$ is shown in the inset of Fig.~2.

In what follows, however, we focus on an alternative approach based on particle trajectories.  
This method leads to an exact analytical solution valid for any waiting-time distribution $\psi(t)$ of the dichotomous fluctuations, thus extending the results beyond the Poissonian case.
 \section{Trajectory approach: general exact solution}
In fact, from Eq. (\ref{electr4})  and from $\omega(t)$ (i.e.$B(t)$) being constant  between switches at times $t_k$ and equal to $\omega_0\pm\omega_1\equiv 2\omega_{\pm}$,
it descends for $u(t)$: 
\BEQ
\label{uss}
u^{\pm}(t) =A^{\pm}_k  e^{i \omega_{\pm} (t-t_k)}+B^{\pm}_k e^{-i \omega_{\pm} (t-t_k)}\;\; , \;\; t_k<t<t_{k+1}\\
\EEQ
where  $\pm$  when added  as superscript  (e.g. to $u$ or $A_k$,$B_k$ in (\ref{uss})) indicates the state of the magnetic field at time $t=0$, (i.e.  the value $\omega(t=0)= \omega_0 \pm \omega_1$)   while  as subscript it refers to the value taken by the field before a switch at  generic time $t_n>0$. 
The  continuity of $u$ and $\dot{u}$ at the changes of the magnetic field  imposes the following conditions on the coefficients $A$ and $B$:
\BEA
\nn A^{\pm}_{2n}&=&A^{\pm}_{2n-1} e^{i \omega_{\mp}\tau_{2n}}W_{\pm}+B^{\pm}_{2n-1} e^{-i \omega_{\mp}\tau_{2n}}\bar{W}_{\pm}\\
\nn B^{\pm}_{2n}&=&A^{\pm}_{2n-1} e^{i  \omega_{\mp}\tau_{2n}}\bar{W}_{\pm} +B^{\pm}_{2n-1} e^{-i \omega_{\mp}\tau_{2n}} W_{\pm}
\EEA
where  $\tau_{2n}=t_{2n}-t_{2n-1}$ and 
we have also introduced the following coefficients
\BEQ
W_{\pm}=\frac{1}{2} \left(1+\frac{\omega_{\pm}}{\omega_{\mp}} \right) \;\; , \;\; \bar{W}_{\pm}=\frac{1}{2} \left(1-\frac{\omega_{\pm}}{\omega_{\mp}} \right).
\EEQ
The  coefficients $A^{\pm}_n, B^{\pm}_n$  can now be calculated recursively, noticing that, for ${\bf A^{\pm}_n}=(A^{\pm}_n, B^{\pm}_n)$ then
\BEQ
{\bf A}^{\pm}_{2n+1}= M_{\pm}(\tau_{2n+1}) {\bf A}^{\pm}_{2n} \;\; , \; \; {\bf A}^{\pm}_{2n}= M_{\mp}(\tau_{2n}) {\bf A}^{\pm}_{2n-1}
\EEQ

with
\begin{gather}
M_{\pm}(\tau)=\begin{bmatrix}
    \exp(i \omega_{\pm} \tau)W_{\pm}       & \exp(-i \omega_{\pm} \tau)\bar{W}_{\pm}\\
    \exp(i \omega_{\pm} \tau)  \bar{W}_{\pm}   &\exp(-i \omega_\pm \tau) W_{\pm}\\
\end{bmatrix}
\end{gather}

Therefore iteratively 
\BEA
\label{iter}
{\bf A}^{\pm}_{2n+1}&=& M_{\pm}(\tau_{2n+1}) M_{\mp}(\tau_{2n}) {\bf A}^{\pm}_{2n-1}\\
\nonumber &=& M_{\pm}(\tau_{2n+1})M_{\mp}(\tau_{2n})\cdots   M_{\mp}(\tau_{2})M_{\pm}(\tau_{1}){\bf A}^{\pm}_{0},
\EEA
where ${\bf A}^{\pm}_{0}$ is fixed by the initial conditions. 
Since   $ r^2(t)=|u(t)|^2$  we can then evaluate    $\langle r^2(t)\rangle$ from  $\langle |u(t)|^2 \rangle$,
and therefore, averaging over the initial value of the field, 
\BEQ \label{r2}
\langle r^2(t)\rangle=\langle |u(t)|^2\rangle=\frac{\langle |u^{+}(t)|^2\rangle+\langle |u^{-}(t)|^2 \rangle}{2} 
\EEQ
In order to evaluate $|u(t)|^2$, from Eq.(\ref{uss})  we get
\BEQ
\label{modu}
|u_n^{\pm}(t)|^2 =|A^{\pm}_{n}|^2 +  |B^{\pm}_{n}|^2 +A^{\pm}_n B^{\pm *}_n e^{2i \omega_{\pm}(t-t_n)}+ c.c.  
\EEQ
which  can  be rewritten as:
\BEQ
\label{equn}
|u_n(t)|^2={\bf A^{\dagger}_n}(t_n)\cdot\sigma_{\pm}(t-t_n) {\bf A_n}(t_n)
\EEQ
with
\BEQ
\sigma_{\pm}(t)=
  \begin{bmatrix}
    1 & e^{2 i\omega_{\pm} t}  \\
    e^{-2 i\omega_{\pm} t}& 1
  \end{bmatrix}
\EEQ
where we have now dropped the $\pm$ superscript indicating the initial value of the field (and we'll also drop  the $\pm$ subscript  in the matrices  $M,\sigma$ indicating the value of the field before each switch) to lighten notation.
 \begin{figure}[h]
\label{Figa}
\centering
\includegraphics[width=.475 \textwidth] {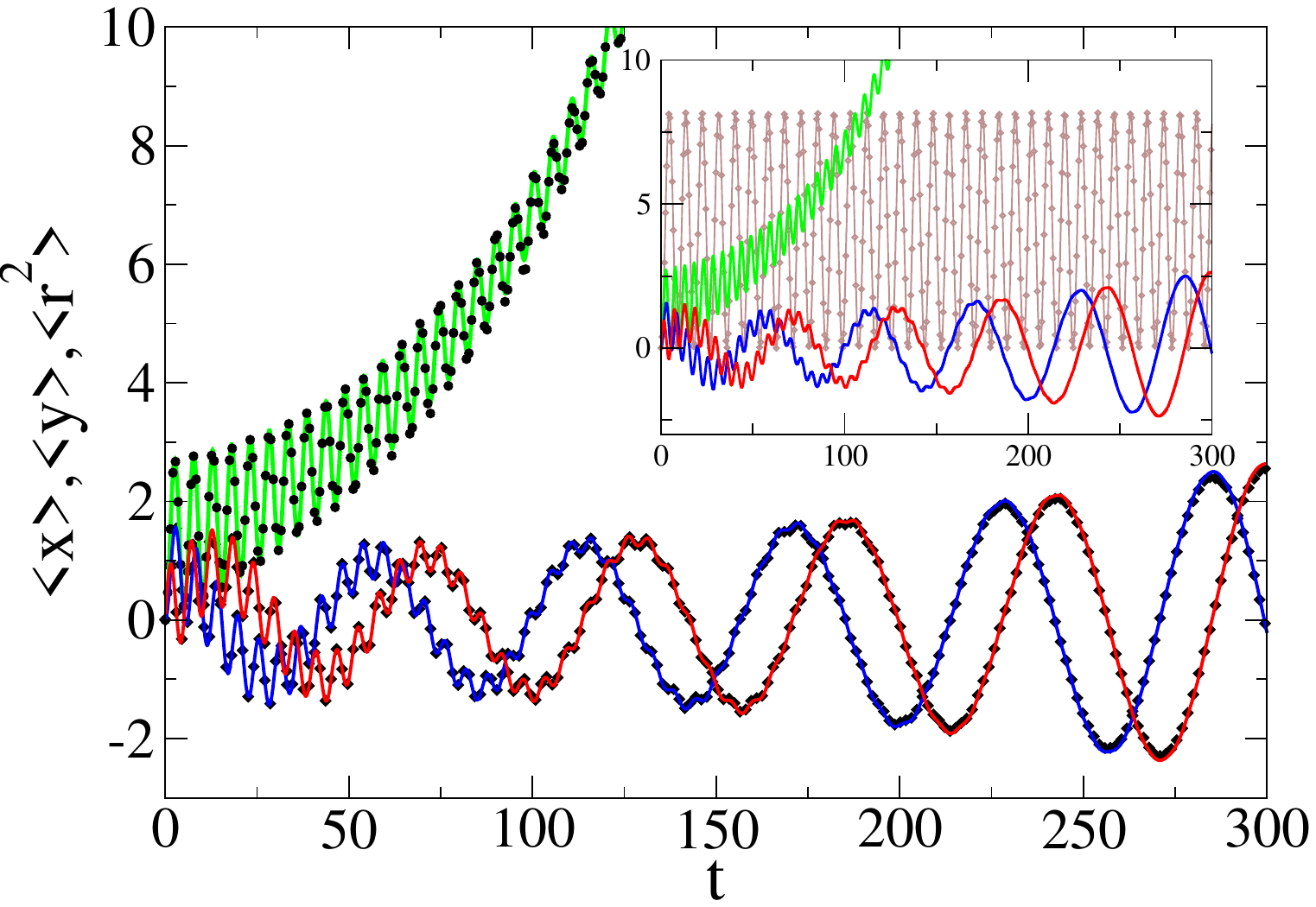} 
\caption{Diffusion of a charged particle in a fluctuating dichotomous magnetic field $B(t)$  in the Poissonian case. The green line is  $\langle r^2 \rangle$  from solution (\ref{exactT}),  blue  and  red line are $\langle x \rangle$, $\langle y \rangle$  from solution  (\ref{quarto2}).
Black diamonds, triangles and circles are the respective simulated results, averaged over 50k runs,  which show perfect agreement. 
Parameters  are $v_0=1,\gamma=20,\,\omega _0 = 1,\,\omega _1= 0.7$ and initial conditions $x(0)=y(0)=\dot{ x}(0)=0$ and $\dot{y}(0)=v_0$ corresponding to ${\bf A}^{\pm}_{0}=\frac{1}{2}(\frac{1}{\omega_{\pm}},-\frac{1}{|\omega_{\pm}|}) $.
Inset:  corresponding solutions  from the density approach Eqs.(\ref{mauror2}), (\ref{quartoZ}). Setting  $\omega_0$=0 in (\ref{mauror2}) and (\ref{exactT}) (brown line) coincides with prediction of Eq. (\ref{confino})  and with simulation (triangles), showing confinement at any time.}
\end{figure}
 In  the case of a trajectory with just  $n=2$  switches of the field at $t_1$ and $t_2$  we have from Eqs. (\ref{equn}) and (\ref{iter}):
\BEA
&&|{\bf u}_2(t)|^2= \sum_i {\bf u^*}_2^i  {\bf u}_2^i ={\bf A^{\dagger}_2}\cdot\sigma(t-t_2) {\bf A_2}\\
\nn &&=\sum_{i,j,k_1,k_2,k_3,k_4} {\bf A^{\dagger}}_{0}(k_2) M^{\dagger}_{k_2,k_1}(\tau_1) M^{\dagger}_{k_1,i}(\tau_2)\sigma_{ij}(t-t_2) \\
\nonumber &&M_{j,k_3}(\tau_2)  M_{k_3,k_4}(\tau_1) {\bf A}_{0}(k_4)
\EEA
Now let ${\bf \tilde{A}_0=A_0^{\dagger}\otimes A_0}$  the  tensor product of the two 2D-vectors $A^{\dagger}_0$ and $A_0$  which can be viewed as the 4D vector  $(A_0^*(1) A_0(1), A_0^*(1)A_0(2),A_0^*(2) A_0(1),A_0^*(2)A_0(2))$.
Analogously let us consider the 4x4 matrix $\tilde{M}(\tau)=M^{\dagger}(\tau)\otimes M(\tau)$ obtained as tensor product 
we can rewrite
\BEQ\label{smart}
|{\bf u}_2(t)|^2={\bf \Sigma}(t-t_2,\tau_2) \cdot \tilde{M}(\tau_1) {\bf \tilde{A}_0} 
\EEQ
where ${\bf \Sigma}$ in (\ref{smart}) is the 4-components vector $ (\Sigma_{1,1},\Sigma_{1,2},\Sigma_{2,1},\Sigma_{2,2})$ with $\Sigma$ the $2\times 2$ matrix
\BEQ
\Sigma(t-t_2,\tau_2)=M^{\dagger}(\tau_2)\sigma(t-t_2)  M(\tau_2) 
\EEQ
We can then  average Eq. (\ref{smart})  over all possible trajectories with 2 switches:
\begin{align} \label{smart1}
&\langle |{\bf u}_2(t)|^2 \rangle =\\
\nonumber&\int_{t_1}^t dt_2\int_0^t dt_1 \Psi(t-t_2) \psi(\tau_2){\bf \Sigma}(t-t_2,\tau_2)\cdot  \tilde{M}(\tau_1) \psi(\tau_1){\bf \tilde{A}_0} 
\end{align}
by including
$\psi(t)$ 
  and  the corresponding survival probability $\Psi(t)=\int_t^{\infty}\psi(t')dt'$, and integrating over all the possible occurrences of two switches  in $(0,t)$.

This result can be generalised to $n$ field switches so that  the final expression for $\langle r^2\rangle=\langle|u|^2\rangle$ is:
\BEA\label{smartF}
&&\langle r^2(t) \rangle =\sum_{n=1}^{\infty} \int_{t_{n-1}}^t dt_n\int_{t_{n-2}}^t dt_{n-1}\dots \int_0^t dt_1 \Psi(t-t_n) \;\;\;\;\; \; \\
\nonumber&& \psi(\tau_n){\bf \Sigma}(t-t_n,\tau_n) \cdot \tilde{M}(\tau_{n-1}) \psi(\tau_{n-1})  \cdots \tilde{M}(\tau_1) \psi(\tau_1){\bf \tilde{A}_0} 
\EEA
to which the contribution  $\langle r^2(t) \rangle^{\pm}_0=\Psi(t) {\bf A^{\pm \dagger}_0}\cdot\sigma_{\pm}(t) {\bf A^{\pm}_0}$ from trajectories with no switches of the field has to be added.
 Thanks to its convolution structure (which is preserved due the exponential time dependence of both ${\bf \Sigma}$  and $\tilde{M}$) Eq. ('\ref{smartF}) can be exactly summed in Laplace domain.
If we now reintroduce the dependence on the initial value of the magnetic field infact this leads to
\BEQ
\label{smartLAP}
 \langle r^2(s) \rangle^{\pm}  =\frac{{\bf \tilde{\Sigma}_{\pm}}(s) \cdot  {\it I}  +{\bf \tilde{\Sigma}_{\mp}}(s)\cdot\tilde{M}_{\psi_\pm} (s) } { 1-\tilde{M}_{\psi _\mp} (s)\tilde{M}_{\psi_\pm}(s) } {\bf \tilde{A}^{\pm}_0} +\langle r^2(s) \rangle^{\pm}_0
 \EEQ
where $\tilde{M}_{\psi_\pm} (s)$ is the Laplace transform $\mathcal{L}_s[\psi(t)\tilde{M}_{\pm} (t)]$, ${\bf \tilde{\Sigma}_{\pm}}(s)$=$\mathcal{L}_s\left[\int_0^t d\tau \Psi(t-\tau)\psi(\tau){{\bf\Sigma}_{\pm}}(t-\tau,\tau)\right]$ and$\;$ $\langle r^2(s) \rangle^{\pm}_0$=$\mathcal{L}_s\left[\langle r^2(t) \rangle^{\pm}_0\right].\;$
The two terms
in the fraction on the right-hand side of Eq. (\ref{smartLAP})  are the result of summing up on trajectories that end either with the field pointing in the opposite direction as at the start (first term), and therefore undergo an even number of field switches before the last one, or pointing in the same direction (second term), and therefore with an odd number of switches before the last one.   Eq. (\ref{smartLAP}) can therefore be expressed as a function only of $\mathcal{L}_s[\psi(\tau)]$ (or of its translates),   and then  used to  evaluate $\langle r^2(t)\rangle$ 
  via Eq. (\ref{r2}).
For Poissonian fluctuations of rate $\gamma$  this leads to 
\BEA
\label{exactLap}
\nonumber &&\langle r^2(s)\rangle=2\left((\gamma+s)^2+(\omega_1^2+\omega_0^2)\right)/\left[s \omega_0^4 +s(s^2 + \omega_1^2)\cdot \right.\\
&& \left.  \left( (\gamma+s)^2+\omega_1^2 \right) +4((\gamma+s)^2+s^2)( s -  \frac{\omega_1^2}{\gamma+s})\omega_0^2  \right]
\EEA
which can be inverted exactly leading to 
\BEQ
\label{exactT}
 \langle r(t)^2\rangle=\sum_{i=1}^3 \sum_{\pm} e^{-\frac{1}{2}(\gamma \pm \sqrt{\gamma^2 +4 \lambda_i})t}  P^{\pm}(\gamma,\omega_0, \omega_1,\lambda_i)
 \EEQ
    with $\lambda_i$ the three roots of  the equation: 
    \BEA
    \label{terzo}
     \nonumber       &&   x^3+ 2( \omega_1^2 +  \omega_0^2) x^2  + \left(\gamma^2 (\omega_1^2 + \omega_0^2)  +(  \omega_1^2 -  \omega_0^2) ^2 \right) x 
     \\&&- \gamma^2 \omega_1^2 \omega_0^2=0 
     \EEA
   and
 \begin{align}
  &P^{\pm}(\gamma,\omega_0, \omega_1,\lambda_i)=\\
   \nonumber &\frac{\mp  \gamma^3 + \sqrt{\gamma^2 + 4\lambda_i} \left(   \gamma^2 +   \lambda_i + \omega_1^2 + \omega_0^2\right) - 
    \gamma \left(3 \lambda_i + \omega_1^2 + \omega_0^2 \right)}{\sqrt{\gamma^2 + 4\lambda_i} (\lambda_i - \lambda_{j \neq i}) (\lambda_i - \lambda_{k\neq j \neq i})}
           \end{align}
Importantly, from Eq. (\ref{terzo})  and Descartes's rule ,  when $\omega_0>0$ it  descends the existence of  at least one positive root, which implies always an exponential increase in time of $\langle r^2(t)\rangle$, i.e. charge acceleration.
In fact from Equations (\ref{kicksa})
 and (\ref{kicksb})
it also follows that:
\BEQ
\label{expproof}
\langle \Delta v^2 (t)\rangle= \omega_1^2 \langle r^2 (t)\rangle
\EEQ
meaning that  the average change in kinetic energy after each switch of field is proportional to  the  spatial second moment and therefore will increase exponentially.

Slightly simpler is the derivation of $\langle z \rangle $ and therefore of $\langle x, y \rangle$, starting from trajectories of $u$ (see Appendix ~S2 for details).
Figure 2 shows $\langle r^2 \rangle$,$ \langle x\rangle$ and $\langle y\rangle$ so evaluated, which coincide with the  result from the density approach (inset) and with the numerical simulation.

\subsection{Average energy gain per magnetic field  switch}

One can then calculate the energy gain $\Delta T$  per magnetic field switch. From the variation of kinetic energy after the field switch if follows
\BEA
\label{intermezzo}
\nonumber && \Delta T=\frac{m}{2}\left((v_x+\Delta v_x)^2+(v_y+\Delta v_y)^2\right)-\frac{m}{2}v_x^2-\frac{m}{2}v_y^2 \\&&
=\frac{m}{2}\Delta v^2 +m \Delta v_x v_y+ m\Delta v_y v_x.
\EEA
Now using again  Equations (\ref{kicksa})
 and (\ref{kicksb})
 we can rewrite Eq. (\ref{intermezzo})  as:
 \BEQ
\label{DT}
\Delta T=\frac{m}{2}\Delta v^2 \pm m \omega_1 (v_x y -v_y x)
\EEQ
and,  since the second term on the right-hand side of (\ref{DT}) averages to zero, summing over all trajectories  using Eq. (\ref{expproof}) we get:
 \BEQ
\label{DTmedio}
\langle \Delta T\rangle=\frac{m}{2}\langle\Delta v^2 \rangle= \frac{m}{2} \omega_1^2 \langle r^2 (t)\rangle,
\EEQ
which will increase exponentially in time as stemming from Eq. (\ref{exactT}).
From this derivation, it is clear that the particle receives an impulse every time the magnetic field switches, due to the induced electric field. It is the latter that does the work that causes the change in kinetic energy, since the magnetic field does not do any work.
We can now compare this average energy gain with the one obtained in the classical Fermi formula, they both lead to an exponential increase in time of the energy gain. But while in the Fermi case the
rate in the exponential is proportional to $V^2/c^2$ (the squared ratio of the velocity of the magnetic cloud and the velocity of the particle, see Eq. (9) in  Ref. \cite{Fermi}) which is  small, in our model,  it is a positive value, independent of the particle velocity, that depends only on the magnetic field parameters and (e.g. in the Poisson case)  is of the order of one for $\omega_1\simeq \omega_0$.  This means that in a regime of high speed particles the acceleration in the latter model would be more effective.

\subsection{The symmetric case: $B_0=\omega_0=0$}
Let us  consider the case of  $B_0=\omega_0=0$, for which  $\omega(t)^2=\omega_1^2=q^2B_1^2/m^2$ since $\xi^2=1$. 
 Even though this can appear as a very particular condition, it is very important physically, as
one can re-create  it by adding a  field $B_{ex}=-B_0$.
  Eq. (\ref{electr4}) in such a case  becomes a trivial harmonic equation.
Therefore, 
with  the  initial condition of $x_0=y_0=v_x(0)=0$ and $v_y(0)=v_0$,  it follows
\begin{eqnarray} \label{confino}
	&&\langle r^2 (t)\rangle=\langle |u(t)|^2 \rangle=\frac{4v_0^2}{\omega_1^2} \sin^2 \left[\frac{ \omega_1 t}{2}\right].
\end{eqnarray}
The last result holds  for any type of fluctuations, namely either Poissonian or non-Poissonian,  and shows a permanent confinement (see brown line in the inset in Figure 2).
Again, we remark  that this result is a direct consequence of including the induced electric field. Without this component, in fact, the dynamics of the particle is diffusive again and no confinement occurs (see \cite{aq3}).   

 {
  \begin{figure}[h]
 
     \includegraphics[scale=.28]{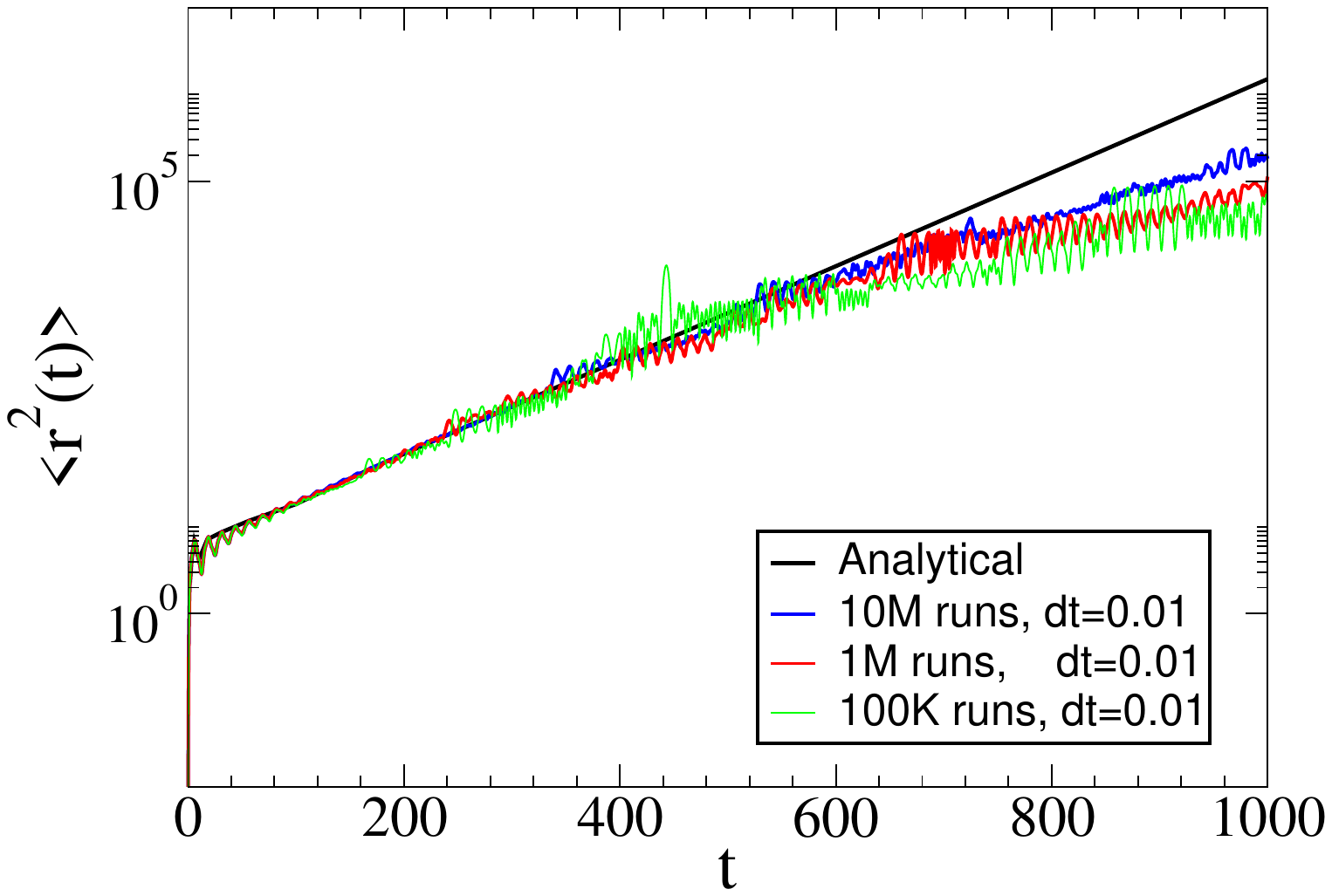} 
  \hspace{.2cm} \includegraphics[scale=.28]{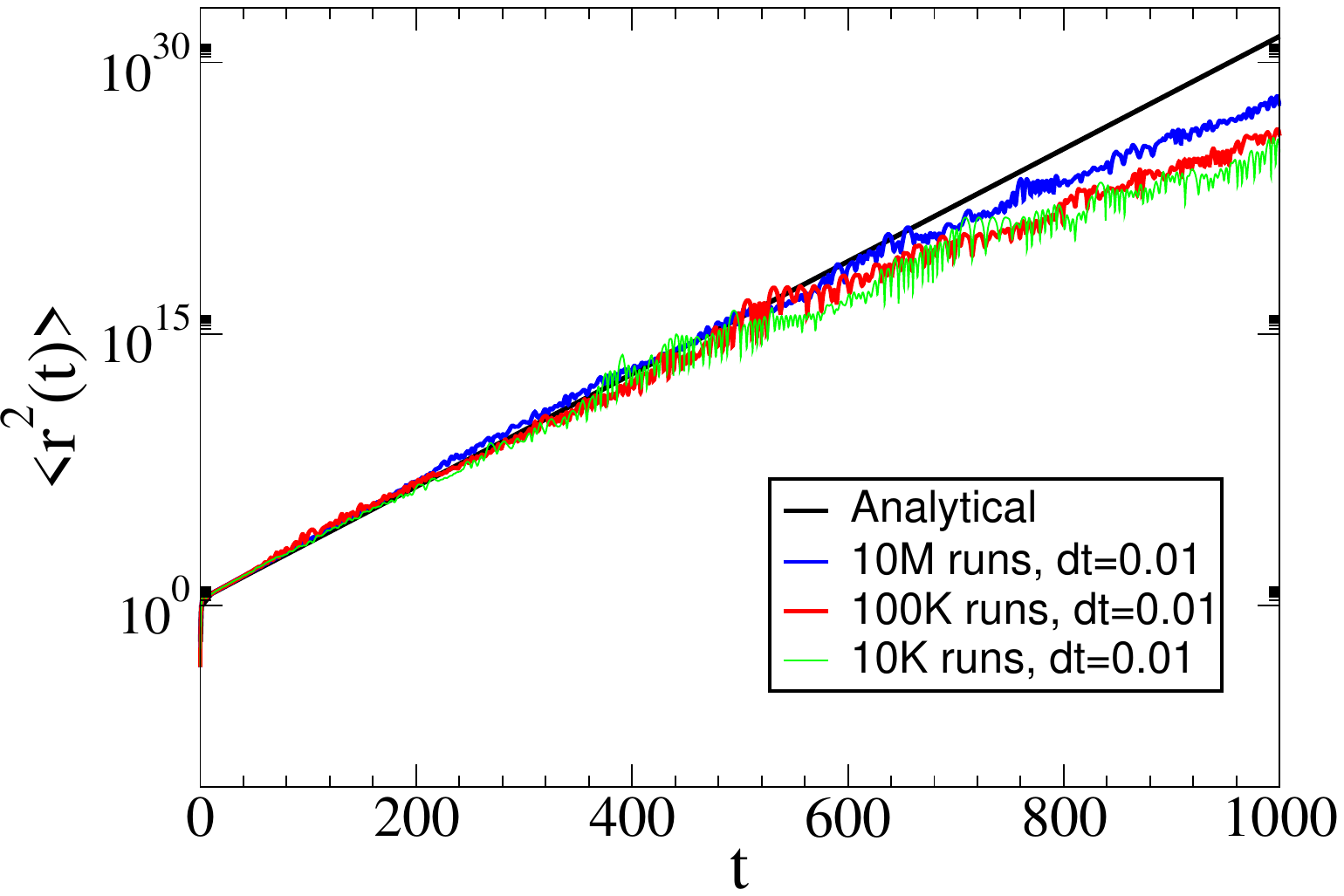} 
    \\ 
\caption{$\langle r^2(t)\rangle$  for the diffusion of a charged particle in a fluctuating magnetic field $B(t)$,  with parameters $\omega_0=1$ and $\omega_1=0.5$  and for field fluctuations with power-law distribution  $\psi(t)_{t\gg1}\sim (t/ \tau)^{-\alpha-1}$. (Left panel)  non-ergodic case of a Mittag-Leffler distribution with parameters   $\tau=10$ and $\alpha=1/2$ (diverging mean time). (Right panel)  Power-law distribution with finite mean time and diverging second moment, as in Eq. (\ref{manneville})  with  $\alpha=3/2$ and $\tau=1$. In both panels, the black lines indicate the result obtained from inverting the analytical solution (\ref{smartLAP}). All other lines:   numerical simulations.
Higher number of runs  show closer match with the analytical solution, with slower convergence for the non-ergodic case.}
    \label{figS3}
  \end{figure}
 }
 
\subsection{Slow fluctuations and ergodicity breaking}
The analytical solution~\eqref{smartLAP} for $\langle r^2(t)\rangle$ can also be used to characterise the diffusion regime ($\omega_0 > 0$) under slow power-law distributed fluctuations and even in condition of ergodicity breaking~\cite{bel,aq2,prl,pnasB}.  
An example is provided by a Mittag–Leffler waiting-time distribution, $\psi(t) = \mathcal{E}_{\alpha}(t/\tau)$, with index $\alpha < 1$, which has a divergent mean sojourn time and has been extensively studied in the field of anomalous diffusion \cite{met,barka, west,superp} and glassy dynamics \cite{glass1,glass2,glass3}. We consider also the case of slow fluctuations with power-law distribution of the form:
\BEQ
\label{manneville}
\psi(t)=\frac{\alpha}{\tau} \frac{1}{(1+t/\tau)^{\alpha +1}}
\EEQ
which for $2<\alpha<3$ has finite mean sojourn time but diverging second moment. This distribution is often found in the context of turbulence \cite{Manneville,aq4}. Note that  both distributions  have an asymptotic power-law behaviour $\psi(t)\propto (t/\tau)^{-\alpha-1} $for $t \gg \tau$.

Figure~3 compares  for both cases the analytical prediction obtained by inverting Eq.~\eqref{smartLAP}  with the numerical simulations.  
Although the closed-form expressions in both cases are algebraically involved, their Laplace transform (see Eq.~\eqref{denmittag}, Eq.~\eqref{denmanneville}  and Fig. \ref{figS2} in the Appendix) exhibit a single positive pole that can be computed exactly.  
This pole determines the exponential growth rate of $\langle r^2(t) \rangle$,  corresponding to the slopes of the black lines in Fig.~3.

Figure 3 shows as well that for these slow fluctuations convergence to the predicted value is slow with increasing number of trajectories, slower in the non-ergodic case. This is due to the fact that many runs get stuck with the same field value for long times, not sampling enough phase space.

Crucially, regardless of the statistics of the fluctuations, any deviation from $B_0 = 0$ leads to the emergence of an exponential diffusion regime, signalling the loss of confinement.

\section{The periodic case}

 \begin{figure}[h]
 \centering  
  \includegraphics[width=0.45\textwidth]{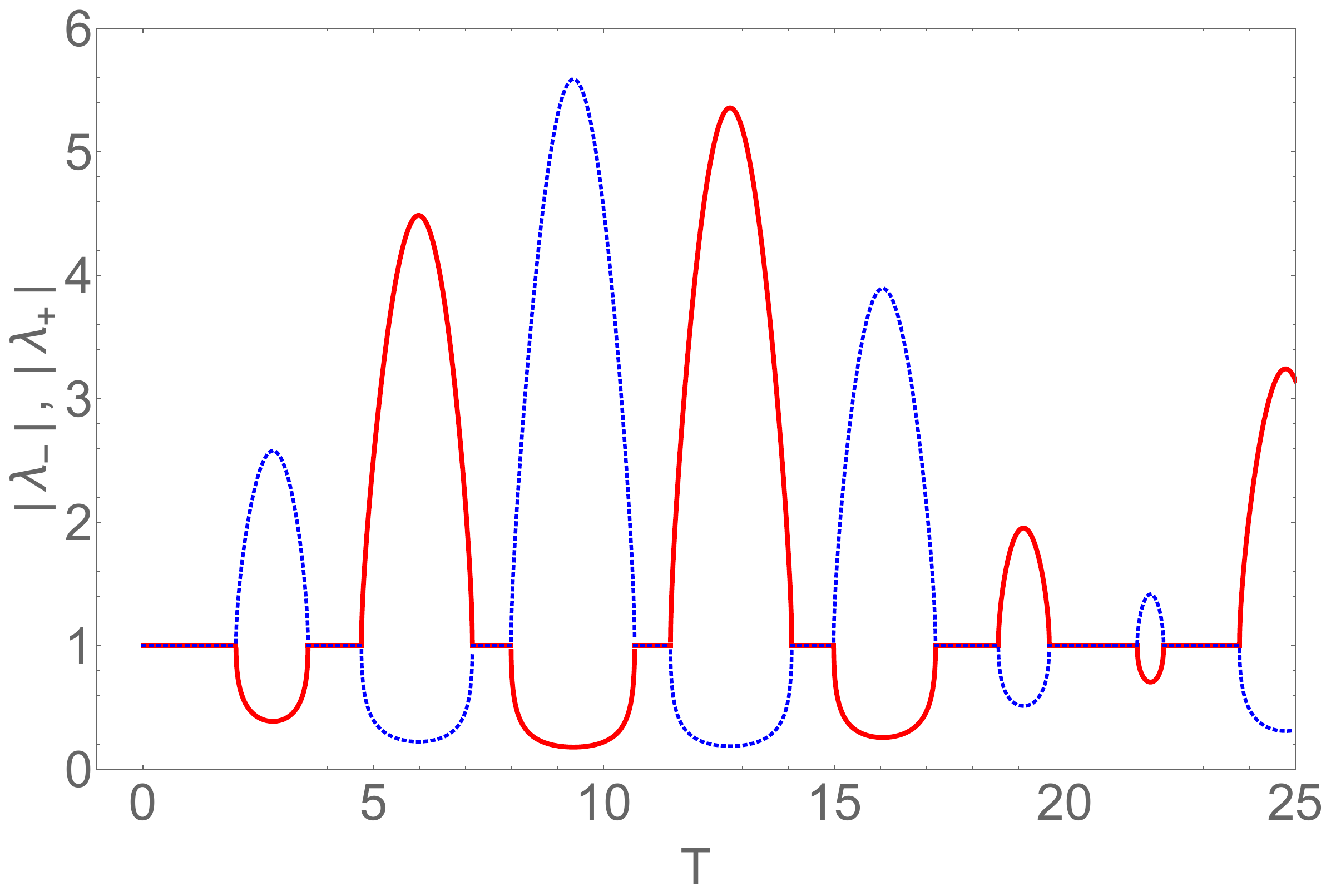}
\caption{Modulus of eigenvalues $\lambda_-$ (red line) and $\lambda_+$ (blue dotted line) of $G_{\pm}=M_{\mp}(T)M_{\pm}(T) $ for the periodic case as a function of period $T$ for $\omega_0=1$ and $\omega_1=0.7$.  For most values of  $T$ it is $|\lambda_{\pm}|=1$, corresponding to confined motion, but for T within  'resonance bands' (e.g. for  $2<T<3.5$ or $4.9<T<7.1$),  one of  $|\lambda_{\pm}|$ is  larger than one, i.e.  resonance occurs between the orbital motion and the field, the charge is accelerated.}%
\end{figure} 
The mechanism underlying this behaviour can be further elucidated within the trajectory formalism by examining the case of strictly periodic field reversals, where the magnetic field switches at fixed intervals $T$.  
In this setting, the dynamics is governed by the eigenvalues of the product matrix $M_{\mp}(T)M_{\pm}(T)$, evaluated at $T$, as derived from Eqs.~\eqref{equn} and~\eqref{iter}.  

For most values of $T$ (see Fig.~4 and Appendix~S1 for details), the particle remains confined.  
However, for specific values of $T$—the so-called 'resonance bands'—the switching frequency of the field resonates with the particle’s natural orbital motion, resulting in sustained acceleration.  
In the stochastic case, random fluctuations are always able to excite these resonance bands, as long as their spectrum is non-zero for frequencies  corresponding to the time scales falling within the bands, thus providing a clear explanation for the robustness of the 'noise-induced resonant acceleration' mechanism.  
Once a finite background field is present ($\omega_0 > 0$), confinement is broken and the particle is driven into an accelerated diffusion regime.


\section{Conclusions}
We have presented an exact analytical solution for the dynamics of a charged particle diffusing in a fluctuating, intermittent magnetic field.  
Using a trajectory-based formalism, we derived a closed solution for the diffusion process, valid for arbitrary waiting-time distributions of the magnetic fluctuations.  
Our results show that, as soon as a finite background field is present ($\omega_0 > 0$), the particle’s motion undergoes a sharp transition: from confinement in the absence of a mean field to a hyper-ballistic regime characterised by exponential growth of the mean-square displacement.  
Remarkably, this accelerated diffusion persists even when the fluctuations follow heavy-tailed power-law statistics.

We interpreted this behaviour in terms of resonance bands that appear in the periodic case, demonstrating that stochastic perturbations can sustain and even enhance these resonant mechanisms.  
This highlights the robustness of 'noise-induced resonant acceleration', which should be regarded as a key ingredient whenever intermittency is present in magnetic fields.

Future extensions of this work could include the addition of a radiative-loss term, allowing an estimate of the resulting energy spectrum.  
Moreover, incorporating collective plasma effects would enable the study of noise-induced heating and its potential implications for space and laboratory plasmas.




\pagebreak
\widetext
\begin{center}
\textbf{\large  Appendix: Supplementary Information\\
  Noise-induced resonant acceleration of a  charge  in an intermittent magnetic field: an exact framework for ergodic and non-ergodic fluctuations}
\end{center}
\setcounter{equation}{0}
\setcounter{figure}{0}
\setcounter{table}{0}
\setcounter{page}{1}
\makeatletter
\renewcommand{\theequation}{A\arabic{equation}}
\renewcommand{\thefigure}{A\arabic{figure}}
\renewcommand{\bibnumfmt}[1]{[A#1]}
\renewcommand{\citenumfont}[1]{A#1}

\section{S1: TRAJECTORY APPROACH: Case of periodic dichotomous fluctuations}

For periodic dichotomous fluctuations one can calculate  $r^2(t)$ after $n$ periods directly from Eq. (\ref{equn}) evaluated at $t=t_n=2nT$ with $T$ the fixed switch time of the 
field. This means also that matrix $M$ is evaluated at $\tau=T$, it follows:
\BEQ
|u_{2n}(t=2nT)|^2={\bf A_0}^{\dagger}\left(M_{\pm}^{\dagger}(T)M_{\mp}^{\dagger}(T)\right)^n \cdot \begin{bmatrix}
    1 & 1  \\
    1& 1
  \end{bmatrix} \cdot \left(M_{\pm}(T)M_{\mp}(T)\right)^n {\bf A_0}
    \EEQ
 The  properties of this dynamics are therefore captured by the eigenvalues of the matrices $G_{\pm}=M_{\pm}(T)M_{\mp}(T)$
which are:
\BEA
\label{lambdas}
&&\lambda_{\pm}=\frac{1}{\omega_0^2-\omega_1^2}\Biggl(\omega_0^2 \cos[\omega_0 T] - \omega_1^2 \cos[ \omega_1 T ]  +  \\
\nonumber &&   \pm 2\sqrt{(\omega_1^2 \cos^2[ \omega_1 T/2]- \omega_0^2 \cos^2[ \omega_0 T/2]) (-\omega_1^2 \sin^2[ \omega_1 T/2] +      \omega_0^2 \sin^2[ \omega_0 T/2])} \Biggr)
\EEA
    {
  \begin{figure}[h]
 
  \hspace{-.39cm}   \includegraphics[scale=.25]{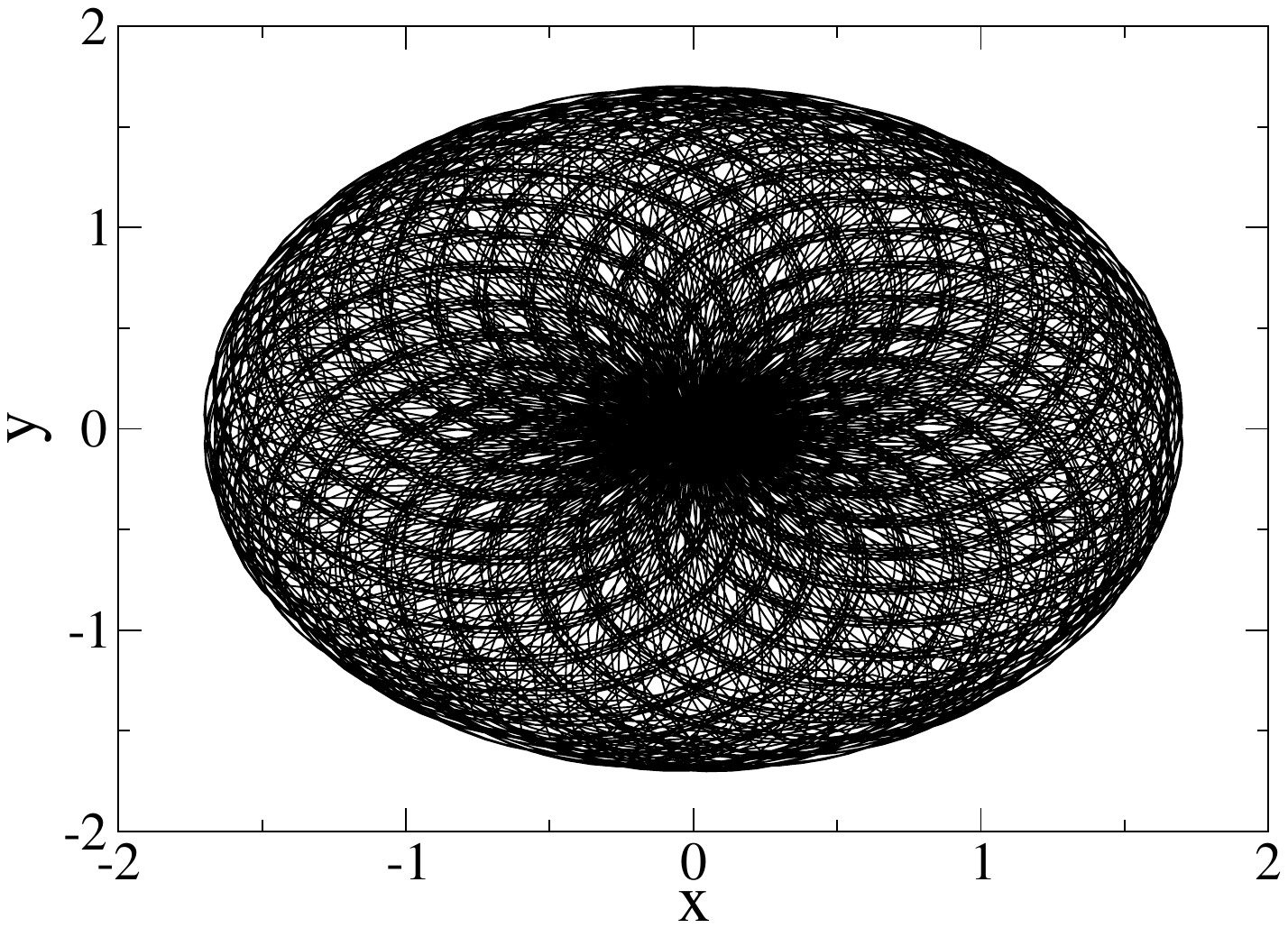} 
  \hspace{.3cm} \includegraphics[scale=.25]{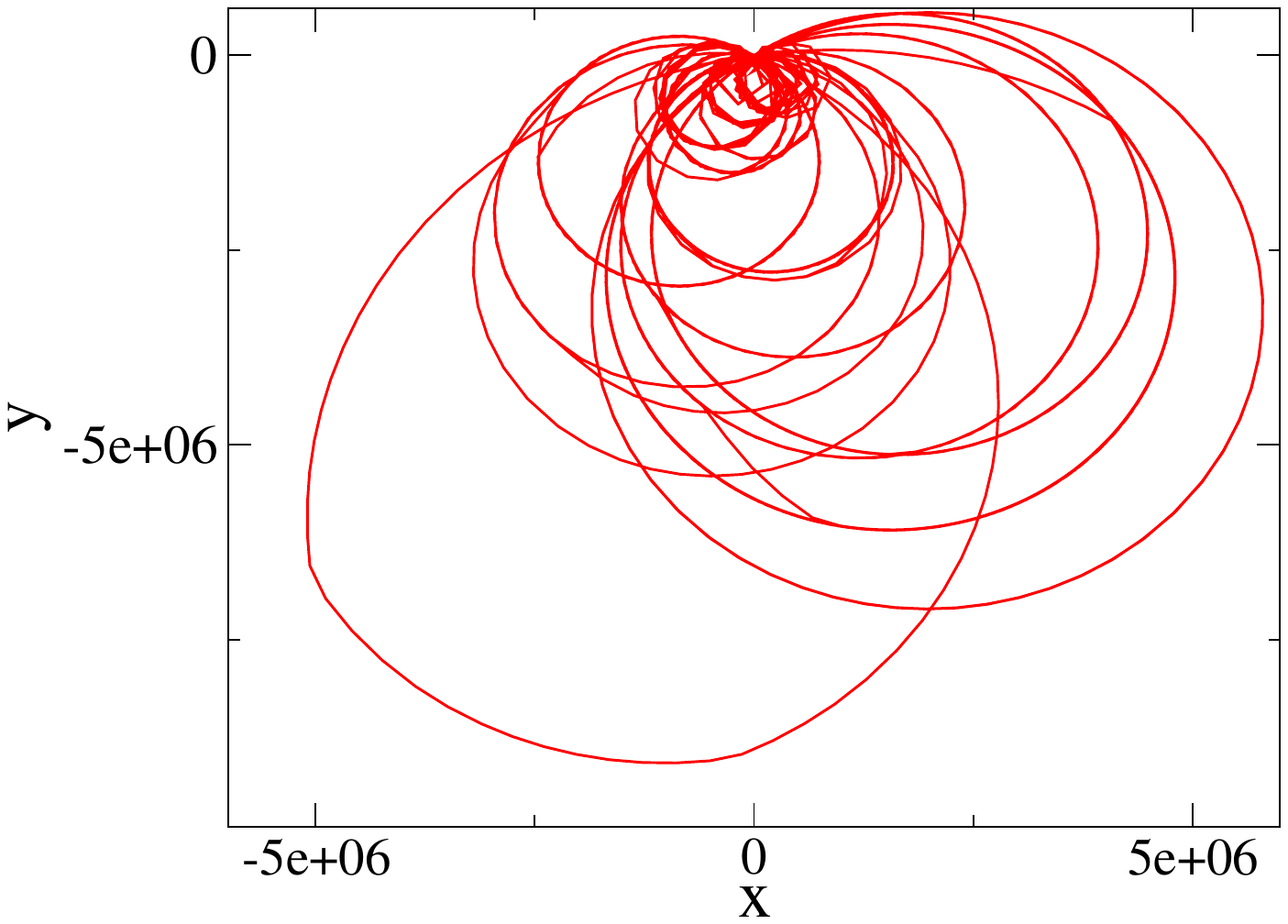} 
    \\ 

    \caption{Examples of trajectories generated with $\omega_0=1$ and $\omega_1=0.7$ at t=10000 for the periodic case with period $T=1$ (left, showing confinement) and $T=3$ (right, showing resonance).}
    \label{figS1}
  \end{figure}
 } 
  For $\omega_0=0$, for instance,  $\lambda_{\pm}=e^{\pm i\omega_1 T}$ and their modulus is one, meaning confined motion.
It is easy to check from (\ref{lambdas}) that this condition of confinement (i.e. $ | \lambda_{\pm}|=1$), applies to every value of $\omega_0, \omega_1$ and $T$ for which the expression within square root is negative.
 When on the contrary, this is positive, one of the eigenvalues necessarily has modulus larger than 1, which leads for $n$ large to an exponentially diverging value for $\langle r^2 \rangle$ which corresponds to resonance.  Example of such conditions are shown in Figure \ref{figS1}, while an example of the resulting 'resonance bands'  (occurring for values of $\omega_0, \omega_1$ and $T$ such that the expression within square root in (\ref{lambdas}) is positive) for different values of the period T are shown in Figure 4 in main text.

\section{S2: TRAJECTORY APPROACH: stochastic fluctuations}

\subsubsection{ Evaluation of $\langle z(t)\rangle$}
In the case of stochastic fluctuations, when the magnetic field switches between its two values  with waiting time distribution $\psi(t)$
one can evaluate  $\langle z(t)\rangle$, from which $\langle x\rangle$ and $ \langle y \rangle$ can be derived as real and imaginary part.
  It is easier to work with variable $u(t)$ which is connected to $z(t)$ via Eq. (\ref{electr3}),
and we know that  for $u$ 
\BEQ
\label{us}
u^{\pm}(t) =A^{\pm}_n  e^{i \omega_{\pm} (t-t_n)}+B^{\pm}_n e^{-i \omega_{\pm} (t-t_n)}\; \;\; \text{for}
\;\;\; t_n<t<t_{n+1} \\
\EEQ
meaning 
\BEQ
\label{us1}
u^{\pm}(t) = \left(  e^{i \omega_{\pm} (t-t_n)},   e^{-i \omega_{\pm} (t-t_n)}\right)\cdot {\bf A^{\pm}_n}
\EEQ
and therefore, using the recursive relations Eqs. (\ref{iter})
\BEQ
\label{us2}
u^{\pm}(t) = \left(  e^{i \omega_{\pm} (t-t_n)},   e^{-i \omega_{\pm} (t-t_n)}\right)\cdot M_{\pm}(\tau_n)\dots M_{\pm}(\tau_1){\bf A^{\pm}_0}
\EEQ
But now from Eq (\ref{electr3}) it is easy to see that  on a single run e.g. with $n$ field switches at times $t_1,\cdots, t_n$ starting and ending with $\xi=+$ 
\BEQ \label{zu}
z(t)=e^{-i\left(\omega_+ (t-t_n)+\omega_-(t_n-t_{n-1}) \cdots+\omega_+t_1\right)} u(t)
\EEQ
and therefore Eq. (\ref{us}) converts for $z$ into:
\BEQ
\label{us3}
 z^{\pm}(t) =\left(  1,   e^{-2i \omega_{\pm} (t-t_n)}\right)\cdot e^{-i \omega_{\pm}\tau_{n}}M_{\pm}(\tau_n)\dots e^{-i \omega_{\pm}\tau_{1}}M_{\pm}(\tau_1){\bf A^{\pm}_0}
\EEQ
and the weight/probability of this run  is $\propto \Psi_+(t-t_n) \psi_-(t_n-t_{n-1}) \cdots\psi_+(t_1)$ where $\psi_{\pm}(\tau) $ is  the  distribution of the  time intervals between magnetic field switches  $\pm\to \mp $   and
\BEQ
\Psi_{\pm}(t)=\int_t^{\infty} \psi_{\pm}(\tau)d\tau
\EEQ
is the corresponding survival probability.
Exploiting  the recursive relations for the cofficients $A^{\pm}_n,B^{\pm}_n$ it follows that the general solution for $\langle z^{\pm}(t) \rangle $ is
\BEQ
\label{rt}
 \langle z^{\pm}(t)\rangle = \sum_{n=0}^{\infty}\int_0^t d\tau \left[\left(\Psi_+(t-\tau), \Psi^*_+ (t-\tau)\right) \cdot {\bf \Phi}_{2n}^{\pm \pm}(\tau) + \left(\Psi_-(t-\tau), \Psi^*_-(t-\tau)\right) \cdot {\bf \Phi}_{2n+1}^{\mp \pm} (\tau)\right]
\EEQ
where  $\Psi^*_{\pm}(\tau)=e^{-2\omega_{\pm} \tau}\Psi_{\pm}(\tau)$ and naming $\tau_n=t_n -t_{n-1}$, 
\BEQ\label{phi}
 {\bf \Phi}_{2n}^{\pm \pm}(\tau)=
\int_0^{\tau} d\tau_{2n}e^{-i \omega_{\mp}\tau_{2n}}\psi_{\mp}(\tau_{2n}) M_{\mp}(\tau_{2n}) \cdots  \int_0^{\tau} d\tau_1e^{-i \omega_{\pm}\tau_{1}}\psi_{\pm}(\tau_1) M_{\pm}(\tau_1) {\bf A}^{\pm}_0
\EEQ
 is the convolution of $2n$  functions $e^{-i \omega_{\pm}\tau}\psi_{\pm}(\tau) M_{\pm}(\tau)$ in correspondence of  $2n$ magnetic field switches in the interval $(0,\tau)$ starting with $\pm$ and ending with same sign $\pm$. We are therefore summing over all the possible realisations, or trajectories, of the fluctuations of the field.
The sum in Eq. (\ref{rt}) can  be evaluated in Laplace transform since, using  the properties of convolution, from (\ref{phi})  it follows:
\BEQ
\label{phis}
\hat{\Phi}_{2n}^{\pm \pm}(s)=\left(M_{\psi_{\mp}}(s+i \omega_{\mp})\right)^n \left(M_{\psi_{\pm}}(s+i \omega_{\pm} )\right)^n{\bf A}^{\pm}_0 \;\; ,\;\;\; \hat{\Phi}_{2n+1}^{\pm \pm}(s)=M_{\psi_{\pm}}(s+i \omega_{\pm} )  \hat{\Phi}_{2n}^{\pm \pm}(s)
\EEQ
where $M_{\psi_{\pm}}(s)=\mathcal{L}_s[\psi_{\pm}(\tau) M_{\pm}(\tau)]$.
This  leads to a geometric series in (\ref{rt}) in Laplace domain that can be evaluated exactly, from the Laplace Transform of  (\ref{rt}):
\BEQ
\langle \hat{z}^{\pm}(s)\rangle = \nonumber \sum_{n=0}^{\infty} \left[\left(\hat{\Psi}_{\pm}(s), \hat{\Psi}^*_{\pm}(s) \right) \cdot {\bf \hat{\Phi}}_{2n}^{\pm \pm}(s) +\left(\hat{\Psi}_{\mp}(s), \hat{\Psi^*}_{\mp}(s) \right) \cdot {\bf \hat{\Phi}}_{2n+1}^{\mp \pm}(s)\right]
\EEQ

it follows via (\ref{phis}):
\BEA
\langle \hat{z}^{\pm}(s)\rangle &=&\left(\hat{\Psi}_{\pm}(s), \hat{\Psi}^*_{\pm}(s)\right) \cdot \frac{1}{1 -M_{\psi_{\mp}}(s+i \omega_{\mp})M_{\psi_{\pm}}(s+i \omega_{\pm})}{\bf A}^{\pm}_0 +\\
\nonumber && \left(\hat{\Psi}_{\mp}(s), \hat{\Psi}^*_{\mp}(s) \right) \cdot \frac{M_{\psi_{\pm}}(s+i \omega_{\pm} ) }{1 -M_{\psi_{\mp}}(s+i \omega_{\mp})M_{\psi_{\pm}}(s+i \omega_{\pm})}{\bf A}^{\pm}_0
\EEA
and the final value for $\langle z \rangle$ can be obtained from inverting:
\BEQ
\langle \hat{z}(s)\rangle=\frac{\langle \hat{z}^{+}(s)\rangle+\langle \hat{z}^{-}(s)\rangle}{2}
\EEQ
with $\langle x \rangle, \langle y \rangle$  then following as  real and imaginary part of $\langle z \rangle$ respectively.
In the Poissonian case of exponential fluctuations, we assume same distribution in each $\pm$ state, i.e.   $\psi_{\pm}(t)=\psi(t)=\frac{\gamma}{2} e^{-\frac{\gamma}{2}  t}$, where we use $\gamma/2$ since in this derivation we have assumed alternating sign in the fluctuations (it can be shown that  the condition of alternating signs maps into the condition of random signs, which is assumed in the density approach,  with distribution $\psi(t)=\gamma e^{-\gamma  t}$,  see also \cite{zumofenklafter}).
  For $\omega_0<\omega_1$  this leads to 
   \BEQ
\langle \hat{z}(s)\rangle=\frac{ \omega_1 
   (s+\gamma/2)}{ s^4  + 2 s^3 (i \omega_0 + \gamma) +  s^2 (\omega_1^2 - \omega_0^2 + 3 \omega_1  \gamma +  \gamma^2)+  s \gamma (\omega_1^2 -\omega_0^2 + i \omega_0  \gamma) + 
   \omega_1^2 \gamma^2/4 
   }
   \EEQ
   for which the inverse Laplace transform is:
      \BEQ
\langle z(t)\rangle= \omega_1 
    \sum_{i=1}^4 \frac{e^{\lambda_i t} (2\lambda_i+ \gamma)} {(\lambda_i - \lambda_{j \neq i}) (\lambda_i - \lambda_{k\neq j \neq i})(\lambda_i - \lambda_{m \neq k\neq j \neq i})}
    \EEQ
    with $\lambda_i$ the four roots of  the equation: 
    \BEQ
    \label{quartoB}
         x^4+ 2( \gamma+ i \omega_0) x^3 +( \omega_1^2 - 
    \omega_0^2 + 3 i \omega_0 \gamma + 
    \gamma^2) x^2 + \gamma( \omega_1^2  - 
     \omega_0^2  + i \omega_0 \gamma)x +\omega_1^2\gamma^2/4=0
     \EEQ
    while for $\omega_0>\omega_1$        one gets
           \BEQ
\langle \hat{z}(s)\rangle=\frac{i (s + \gamma) (s + i \omega_1 + 
    \gamma)}{ (s^4  + 2 s^3 (i \omega_0 + \gamma) +  s^2 (\omega_1^2 - \omega_0^2 + 3 \omega_0  \gamma +  \gamma^2)+  s \gamma (\omega_1^2 -\omega_0^2 + i \omega_0  \gamma) + 
   \omega_1^2 \gamma^2 /4)
   }
   \EEQ
      for which the inverse Laplace transform is:
      \BEQ
        \label{quarto2}
\langle z(t)\rangle= i  \sum_{i=1}^4 
 \frac{e^{\lambda_i t}(\lambda_i^2 +\lambda_i (2 i \gamma-\omega_0 ) + i \gamma^2-\omega_0 \gamma)} {  (\lambda_i - \lambda_{j \neq i}) (\lambda_i - \lambda_{k\neq j \neq i})(\lambda_i - \lambda_{m \neq k\neq j \neq i}) }
    \EEQ    
     with $\lambda_i$ the four roots of  the same equation (\ref{quartoB}).
     
                 
     \subsubsection{Evaluation of $\langle r^2(t)\rangle$: the case of non-ergodic and ergodic power law fluctuations }
     We can use general solution  Eq. (\ref{smartLAP})  to evaluate  $\langle r^2(t)\rangle$ also for non-Poissonian fluctuations, just be replacing the waiting times distribution $\psi(t)$ with the appropriate choise.
     For a Mittag-Leffler distribution,  which asymptotically behaves as  $\psi_{t \gg \tau}(t)\sim (t/\tau)^{-\alpha-1}$ the Laplace Transform is
     \BEQ
     \hat{\psi}(s)=\frac{1}{1+ (s \tau)^\alpha}
     \EEQ
        Replacing this into Eq. (\ref{smartLAP}) gives a complicated expression which can be inverted numerically and is shown in Figure 3 in main text and analysed analytically by studying the poles of the resulting function in Laplace space, we include for completeness the expression for $\alpha=1/2, \omega_0=1,\omega_1=1/2, \tau=10$ used in the text.
        The mean square radius  can be written as:
\BEQ
\langle r^2(s)\rangle=\frac{N(s)}{D(s)}
\EEQ
with $N(s)$ a long algebraic function we do not include  while we include $D(s)$ because determines the asymptotic behaviour in time of the mean square radius.

It is:

{\footnotesize
\begin{equation}
\label{denmittag}
\begin{split}
&D(s)=s (1 + 4 s^2) (9 + 4 s^2) (1 + \sqrt{s \tau}) (2 + \sqrt{2} \sqrt{(-i + 2 s) \tau}) (2 + \sqrt{2} \sqrt{(i + 2 s) \tau}) (2 + \sqrt{2} \sqrt{(-3i + 2 s) \tau}) (2 + \sqrt{2} \sqrt{(3i + 2 s) \tau})\\
&\times  \bigg(2 \sqrt{9 + 4 s^2} \sqrt{2 + 8 s^2} \tau^2 (-8 + 9 \sqrt{s \tau}) + 4 \tau \bigg(17 \sqrt{2 + 8 s^2} \sqrt{s \tau} + 17 \sqrt{18 + 8 s^2} \sqrt{s \tau} - 4 \sqrt{9 + 4 s^2} \sqrt{(-i + 2 s) \tau} \\
&+ 13 \sqrt{9 + 4 s^2} \sqrt{s \tau} \sqrt{(-i + 2 s) \tau} - 4 \sqrt{9 + 4 s^2} \sqrt{(i + 2 s) \tau} + 13 \sqrt{9 + 4 s^2} \sqrt{s \tau} \sqrt{(i + 2 s) \tau} 
- 4 \sqrt{1 + 4 s^2} \sqrt{(-3i + 2 s) \tau} \\
&+ 13 \sqrt{1 + 4 s^2} \sqrt{s \tau} \sqrt{(-3i + 2 s) \tau} - 4 \sqrt{1 + 4 s^2} \sqrt{(3i + 2 s) \tau} + 13 \sqrt{1 + 4 s^2} \sqrt{s \tau} \sqrt{(3i + 2 s) \tau}\bigg)  + 4 \sqrt{2} \sqrt{s \tau} \bigg(\sqrt{(-i + 2 s) \tau} \sqrt{(-3i + 2 s) \tau} \\
&+ 16 \sqrt{(i + 2 s) \tau} \sqrt{(-3i + 2 s) \tau} + 16 \sqrt{(-i + 2 s) \tau} \sqrt{(3i + 2 s) \tau} + \sqrt{(i + 2 s) \tau} \sqrt{(3i + 2 s) \tau}\bigg) + s \tau \bigg(9 \sqrt{9 + 4 s^2} \sqrt{2 + 8 s^2} \tau^2 \\
&+ 18 \tau \bigg(\sqrt{2 + 8 s^2} + \sqrt{18 + 8 s^2} + \sqrt{9 + 4 s^2} \sqrt{(-i + 2 s) \tau} + \sqrt{9 + 4 s^2} \sqrt{(i + 2 s) \tau} + \sqrt{1 + 4 s^2} \sqrt{(-3i + 2 s) \tau} + \sqrt{1 + 4 s^2} \sqrt{(3i + 2 s) \tau}\bigg) \\
-& 2 \bigg(16 \sqrt{(-i + 2 s) \tau} + 16 \sqrt{(i + 2 s) \tau} + 16 \sqrt{(-3i + 2 s) \tau} + 7 \sqrt{2} \sqrt{(-i + 2 s) \tau} \sqrt{(-3i + 2 s) \tau} 
- 8 \sqrt{2} \sqrt{(i + 2 s) \tau} \sqrt{(-3i + 2 s) \tau} \\ 
&+ 16 \sqrt{(3i + 2 s) \tau}  - 8 \sqrt{2} \sqrt{(-i + 2 s) \tau} \sqrt{(3i + 2 s) \tau} + 7 \sqrt{2} \sqrt{(i + 2 s) \tau} \sqrt{(3i + 2 s) \tau}\bigg)\bigg)\bigg)
\end{split}
\end{equation}
}

we can plot the denominator $D(s)$  which shows a pole at $s\approx0.0125$ (see Fig. \ref{figS2})  which coincides with the slope of the black line in Fig. 3 in main text.

For the case of distribution with  finite first moment  but diverging second moment, we consider the distribution given in Eq. (\ref{manneville}) which has the following Laplace Transform
\BEQ
\hat{\psi}(s)=\frac{\alpha}{\tau} e^{s \tau} E_{\alpha+1}[s]
\EEQ
with $E_{\alpha}(x) $ the ExpIntegralE function. As in main text we choose $\tau=1$ and $\alpha=3/2$.
Using Eq. (\ref{smartLAP}) for the mean square radius we get again:
\BEQ
\langle r^2(s)\rangle=\frac{N(s)}{D(s)}
\EEQ
with now
{\footnotesize
\begin{multline*}
N(s)=4 \left(2+3 e^s \text{E}_{5/2}\left[s\right]\right) 
\bigg(-16 e^{\frac{i}{2}} \bigg(-6 e^{\frac{3 i}{2}} \left(5+4 s^2\right) + e^s (3 i+2 s)^2\text{E}_{5/2}\left[-\frac{3i}{2}+s\right]
+ e^{3 i+s} (3 i-2 s)^2 \text{E}_{5/2}\left[\frac{3i}{2}+s\right]\bigg) \\
+ 12 e^{i+s} \text{E}_{5/2}\left[\frac{ i}{2}+s\right] \bigg(12 e^{\frac{3 i}{2}} (i-2 s)^2 - 32 e^{3 i+s} \left(3-2 i s+4 s^2\right)\text{E}_{5/2}\left[\frac{3i}{2}+s\right]\\
+ e^s\text{E}_{5/2}\left[-\frac{3i}{2}+s\right]\bigg(6-32 i s-8 s^2 + 3 e^{\frac{3 i}{2}+s} \left(-9+20 i s+4 s^2\right) \text{E}_{5/2}\left[\frac{3i}{2}+s\right]\bigg)\bigg) \\
+ 6 e^s \text{E}_{5/2}\left[-\frac{ i}{2}+s\right] \bigg(24 e^{\frac{3 i}{2}} (i+2 s)^2 - 2 e^{3 i+s} \bigg(-6-32 i s+8 s^2 + 9 e^{\frac{i}{2}+s} \left(-3+20 i s+12 s^2\right) \text{E}_{5/2}\left[\frac{ i}{2}+s\right]\bigg) \text{E}_{5/2}\left[\frac{3i}{2}+s\right]\\+ e^s\text{E}_{5/2}\left[-\frac{3i}{2}+s\right]\bigg(-64 \left(3+2 i s+4 s^2\right) + 6 e^{\frac{3 i}{2}+s} \left(-9-20 i s+4 s^2\right)\text{E}_{5/2}\left[\frac{3i}{2}+s\right]
+ 9 e^{\frac{i}{2}+s} \text{E}_{5/2}\left[\frac{ i}{2}+s\right] \\ \times \bigg(6+40 i s-24 s^2 + 5 e^{\frac{3 i}{2}+s} \left(9+20 s^2\right) \text{E}_{5/2}\left[\frac{3i}{2}+s\right]\bigg)\bigg)\bigg) + e^s \text{E}_{5/2}\left[s\right] \bigg(-80 e^{2 i} \left(9+20 s^2\right) \\ +72 e^{\frac{i}{2}+s} \left(-3+20 i s+12 s^2\right)\text{E}_{5/2}\left[-\frac{3i}{2}+s\right]
+ 72 e^{\frac{7 i}{2}+s} \left(-3-20 i s+12 s^2\right)\text{E}_{5/2}\left[\frac{3i}{2}+s\right]- 6 e^{i+s} \text{E}_{5/2}\left[\frac{ i}{2}+s\right]\\ \times \bigg(3 e^s\text{E}_{5/2}\left[-\frac{3i}{2}+s\right]
 \left(6+32 i s-8 s^2+27 e^{\frac{3 i}{2}+s} (i+2 s)^2 \text{E}_{5/2}\left[\frac{3i}{2}+s\right]\right) \\- 4 e^{\frac{3 i}{2}} \bigg(9+20 i s-4 s^2+24 e^{\frac{3 i}{2}+s} \left(3+2 i s+4 s^2\right) \text{E}_{5/2}\left[\frac{3i}{2}+s\right]\bigg)\bigg) 
- 3 e^s \text{E}_{5/2}\left[-\frac{ i}{2}+s\right]\\ \times \bigg(8 e^{\frac{3 i}{2}} \left(-9+20 i s+4 s^2\right) - 6 e^{3 i+s} \bigg(-6+32 i s+8 s^2+3 e^{\frac{i}{2}+s} (3 i+2 s)^2 \text{E}_{5/2}\left[\frac{ i}{2}+s\right]\bigg) \text{E}_{5/2}\left[\frac{3i}{2}+s\right]\\+ 3 e^s\text{E}_{5/2}\left[-\frac{3i}{2}+s\right]\bigg(-64 \left(3-2 i s+4 s^2\right) + 54 e^{\frac{3 i}{2}+s} (i-2 s)^2\text{E}_{5/2}\left[\frac{3i}{2}+s\right]
+ 3 e^{\frac{i}{2}+s} \text{E}_{5/2}\left[\frac{ i}{2}+s\right]\\ \times \bigg(-2 (3 i-2 s)^2 + 27 e^{\frac{3 i}{2}+s} \left(5+4 s^2\right) \text{E}_{5/2}\left[\frac{3i}{2}+s\right]\bigg)\bigg)\bigg)\bigg)\bigg) \hspace{7.7cm}
\end{multline*}
}
and
  
{\footnotesize
\begin{equation}
\label{denmanneville}
\begin{split}
&D(s)=3 s (1+4 s^2) (9+4 s^2) \times \\
&\biggl(\text{E}_{5/2}[s]^2 \biggl(-400 e^{2 i+2 s} + 36 e^{i+4 s} \text{E}_{5/2}\Bigl[\frac{i}{2}+s\Bigr] \Bigl(\text{E}_{5/2}\Bigl[-\frac{3i}{2}+s\Bigr] + 16 e^{3 i} \text{E}_{5/2}\Bigl[\frac{3i}{2}+s\Bigr]\Bigr) \\ 
&\quad - 9 e^{4 s} \text{E}_{5/2}\Bigl[-\frac{i}{2}+s\Bigr] \biggl(-4 e^{3 i}\text{E}_{5/2}\Bigl[\frac{3i}{2}+s\Bigr] + \text{E}_{5/2}\Bigl[-\frac{3i}{2}+s\Bigr] \Bigl(-64 + 81 e^{2 i+2 s} \text{E}_{5/2}\Bigl[\frac{i}{2}+s\Bigr] \text{E}_{5/2}\Bigl[\frac{3i}{2}+s\Bigr]\Bigr)\biggr)\biggr) \\
&\quad + 4 \biggl(16 e^{2 i} - 4 e^{i+2 s} \text{E}_{5/2}\Bigl[\frac{i}{2}+s\Bigr] \Bigl(\text{E}_{5/2}\Bigl[-\frac{3i}{2}+s\Bigr] + 16 e^{3 i} \text{E}_{5/2}\Bigl[\frac{3i}{2}+s\Bigr]\Bigr) \\
&\quad + e^{2 s} \text{E}_{5/2}\Bigl[-\frac{i}{2}+s\Bigr] \biggl(-4 e^{3 i}\text{E}_{5/2}\Bigl[\frac{3i}{2}+s\Bigr] + \text{E}_{5/2}\Bigl[-\frac{3i}{2}+s\Bigr] \Bigl(-64 + 225 e^{2 i+2 s} \text{E}_{5/2}\Bigl[\frac{i}{2}+s\Bigr] \text{E}_{5/2}\Bigl[\frac{3i}{2}+s\Bigr]\Bigr)\biggr)\biggr) \\
&\quad \times \text{E}_{5/2}\Bigl[-\frac{3i}{2}+s\Bigr] - 32 e^{\frac{i}{2}+2 s} \text{E}_{5/2}[s] \biggl(-4 \Bigl(\text{E}_{5/2}\Bigl[-\frac{3i}{2}+s\Bigr] + e^{3 i} \text{E}_{5/2}\Bigl[\frac{3i}{2}+s\Bigr]\Bigr) \\
&\quad + e^{2 i} \text{E}_{5/2}\Bigl[\frac{i}{2}+s\Bigr] \Bigl(-4 + 9 e^{2 s} \text{E}_{5/2}\Bigl[-\frac{3i}{2}+s\Bigr] \text{E}_{5/2}\Bigl[\frac{3i}{2}+s\Bigr]\Bigr) + \text{E}_{5/2}\Bigl[-\frac{i}{2}+s\Bigr] \biggl(9 e^{2 s} \text{E}_{5/2}\Bigl[\frac{i}{2}+s\Bigr] \\
&\quad \times \Bigl(\text{E}_{5/2}\Bigl[-\frac{3i}{2}+s\Bigr] + e^{3 i} \text{E}_{5/2}\Bigl[\frac{3i}{2}+s\Bigr]\Bigr) + e^i \Bigl(-4 + 9 e^{2 s} \text{E}_{5/2}\Bigl[-\frac{3i}{2}+s\Bigr] \text{E}_{5/2}\Bigl[\frac{3i}{2}+s\Bigr]\Bigr)\biggr)\biggr)\biggr)
\end{split}
\end{equation}
}

Again one can determine the exponential behaviour of the mean square radius from the poles of the denominator, which show a single pole at about $s\approx 0.071$ which coincides with slope of the black line in Fig. 3 (right panel).
 
 {
  \begin{figure}[h]
 
    \includegraphics[scale=.45]{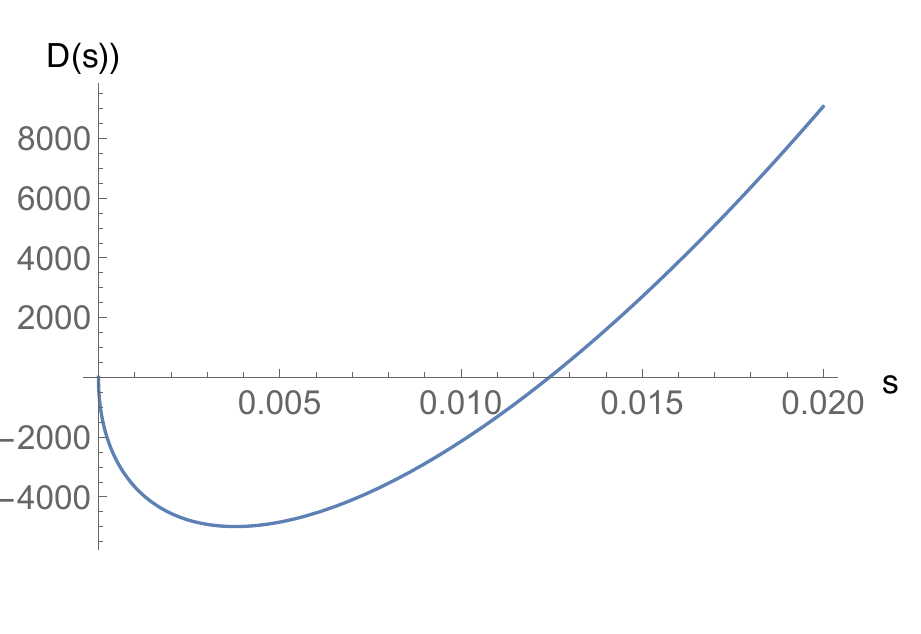} 
  \hspace{1.29cm}
   \includegraphics[scale=.45]{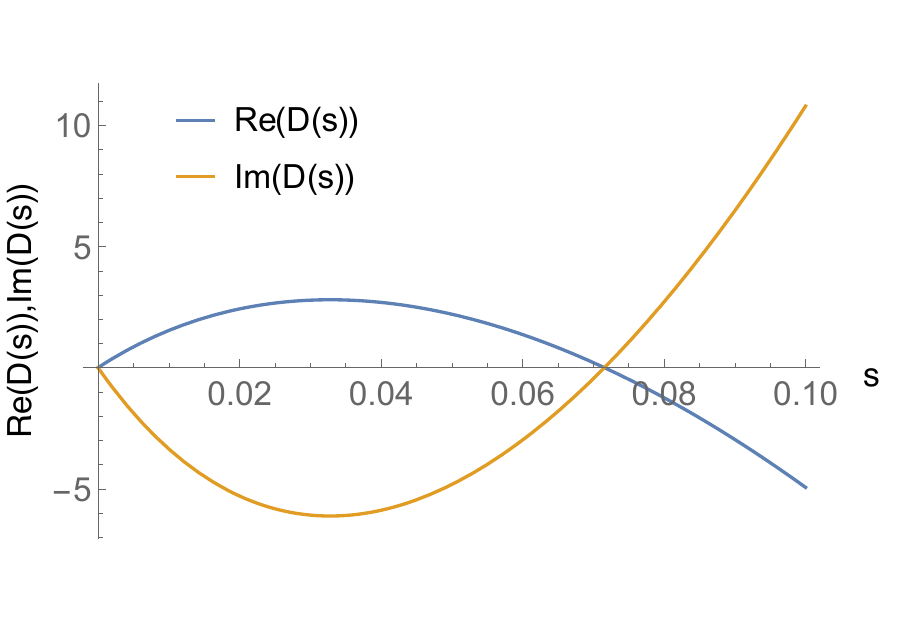} 
    \\ 

    \caption{Denominators for the mean square radius  for $\omega_0=1$ and $\omega_1=0.5$  for the case of Mittag-Leffler function (left, with $\alpha=1/2$ and $\tau=10$)  and power-law distribution with $\alpha=3/2$ and $\tau=1$ (right) used in main text. Both denominators show a single positive pole which determines the exponential rate of the mean radius square, these are the slopes of the black lines in Fig. 3 in main text.}
    \label{figS2}
  \end{figure}
 } 
\vspace{3.275cm}
   \section{S3: DENSITY APPROACH}

             \subsubsection{ Evaluation of $\langle z(t)\rangle$}

Within the density approach, valid only for Poissonian flutctuations, we can evaluate 
$$\langle z\rangle=\langle x\rangle+i\langle y\rangle=U(t),\,\,\langle \omega_1 \dot{z}\rangle=W_p(t),\,\ \langle\omega_1 z\rangle=W(t)$$.
 Taking the average of Eq. (\ref{electr2}) we have
\begin{eqnarray} \label{aver_1}
&&\frac{\partial ^2U}{\partial t^2}+i \omega _0 \frac{\partial U}{\partial t}+i W_p-\frac{1}{2} i \gamma  W=0
\\\label{aver_2}
&&\frac{\partial W}{\partial t}=W_p-\gamma  W
\\\nonumber
&&\frac{\partial W_p}{\partial t}=-\gamma W_p+\langle \omega_1 \ddot{z}\rangle=-\gamma W_p-\Bigr\langle\frac{ i  \dot{\omega}_1 \omega _1 z}{2}+i \omega _1^2 \dot{z}+i \omega _0 \omega _1 \dot{z}\Bigr\rangle=
\\\label{aver_3}
&&-i \omega _1^2 \frac{\partial U}{\partial t}-\left(\gamma +i \omega _0\right) W_p
\end{eqnarray}
where we used that $\langle \dot{\omega}_1 z\rangle=-\gamma W$, $\langle \dot{\omega}_1 \dot{z}\rangle=-\gamma W_p$, and we used Eq.~(\ref{electr2}) for $ \ddot{z}$.  Taking the derivative of Eq.~(\ref{aver_1}) jointly with  Eq.~(\ref{aver_1}) itself, we are able to find $W$ and $W_p$ as a function of $U$ and its derivatives. Taking a further derivative, we end up with
\begin{eqnarray} \nonumber
&&
U^{\prime\prime\prime \prime}(t)+2 \left(\gamma +i \omega _0\right) U^{\prime \prime \prime}(t)+\left(\gamma ^2+3 i \gamma  \omega _0-\omega _0^2+\omega _1^2\right) U''(t)+
\\ \label{correct}
&&\gamma  \left(i \gamma  \omega _0-\omega _0^2+\omega _1^2\right) U'(t)+\frac{1}{4} \gamma ^2 \omega _1^2 U(t)=0
\end{eqnarray}
with initial conditions
 \begin{eqnarray}  \label{incon}
 U(0)=0,\,\,U'(0)=iv_0,\,\,U''(0)=\omega _0 v_0,\,\,U^{\prime \prime \prime}(0)=-i \left(\omega _0^2+\omega _1^2\right)v_0
 \end{eqnarray}
as can be inferred from Eqs. (\ref{aver_1})-(\ref{aver_3}). 
The characteristic equation for Eq.~(\ref{correct}) is
\BEQ
 \label{char_U}
 \lambda^{4}+2 \left(\gamma +i \omega _0\right) \lambda^{3}+\left(\gamma ^2+3 i \gamma  \omega _0-\omega _0^2+\omega _1^2\right) \lambda^2+
\gamma  \left(i \gamma  \omega _0-\omega _0^2+\omega _1^2\right) \lambda+\frac{1}{4} \gamma ^2 \omega _1^2 =0,
\EEQ
(which coincides with Eq. (\ref{quartoB})) and the solution can be found as 
\BEQ
\label{quartoZ}
U(t)=\sum_{i=1}^4 c_i \exp[\lambda_i t]
\EEQ
 with $c_i$ constants determined by conditions (\ref{incon}), and $\lambda_i$ solutions of Eq. (\ref{char_U}).

\subsubsection{ Evaluation of  $\langle r^2(t)\rangle$}
Analogously, we may find the equation for the quantity~$\langle r^2\rangle=\langle x^2\rangle+\langle y^2\rangle$ that describes the diffusion of the particle and is related to the plasma confinement. Setting for sake of compactness~$r_2\equiv \langle r^2\rangle$, then we have\\
\begin{eqnarray} \label{mauror2} 
&&
r^{(6)}_2+3 \gamma r^{(5)}_2+\left[3 \gamma ^2+2 \left(\omega _0^2+\omega _1^2\right)\right]r^{(4)}_2+\gamma  \left[\gamma ^2+4 \left(\omega _0^2+
\omega _1^2\right)\right] r^{(3)}_2 +
\\\label{correct_r2}
\nonumber &&+\left[3 \gamma ^2 \left(\omega _0^2+\omega _1^2\right)+\left(\omega _0^2-\omega _1^2\right){}^2\right]r^{(2)}_2+\gamma  \left[\gamma ^2 \left(\omega _0^2+\omega _1^2\right)+\left(\omega _0^2-\omega _1^2\right){}^2\right]r^{(1)}_2
-\gamma ^2 \omega _0^2 \omega _1^2 r_2=0
\end{eqnarray}
\\
with initial conditions 
\begin{eqnarray} \nonumber
&&
r_2(0)=0,\,\,r^{(1)}_2(0)=0,\,\,r^{(2)}_2(0)=2 v_0^2,\,\,r^{(3)}_2(0)=0,
\\ \nonumber
&&r^{(4)}_2(0)=-2\left(\omega _0^2+\omega _1^2\right)v_0^2,\,\,r^{(5)}_2(0)=0.
\end{eqnarray}

The characteristic equation for Eq.~(\ref{correct_r2}) is\\
\begin{eqnarray} 
&&
\lambda^{6}+3 \gamma \lambda^{5}+\left[3 \gamma ^2+2 \left(\omega _0^2+\omega _1^2\right)\right]\lambda^{4}_2+\gamma  \left[\gamma ^2+4 \left(\omega _0^2+
\omega _1^2\right)\right] \lambda^{3}+
\\\label{caratt_r2}
\nonumber&&\left[3 \gamma ^2 \left(\omega _0^2+\omega _1^2\right)+\left(\omega _0^2-\omega _1^2\right){}^2\right]\lambda^{2}+ \gamma  \left[\gamma ^2 \left(\omega _0^2+\omega _1^2\right)+\left(\omega _0^2-\omega _1^2\right){}^2\right]\lambda^{1}
-\gamma ^2 \omega _0^2 \omega _1^2 =0.
\end{eqnarray}
\\

Both Eqs.~(\ref{correct}) and (\ref{correct_r2}) can be solved analytically, being differential equations with constant coefficients. Being the characteristic equations of the fourth and sixth degree, respectively, the expressions for the constants $\lambda_i$ are not easy to handle. 

Let us focus on the solution of Eq.~(\ref{correct_r2}), describing the diffusion of the particle.
Using Descartes' rule of signs, we deduce that there is a positive root, which implies that, in general, the particle diffuses and $r_2(t)\to\infty$ for $t\to\infty$. The full solution of Eq.~(\ref{caratt_r2}) gives an unmanageable analytical expression. We may find an analytical approximated solution considering the case $\omega _0\ll \omega _1$. Due to the evident symmetry of the polynomial with respect to the exchange of $\omega _0$ and $\omega _1$, the result will also apply to $\omega _1\ll \omega _0$. Setting $\varepsilon\equiv \omega _0/ \omega _1$ we have

\begin{align} 
&\lambda^{6}+3 \gamma \lambda^{5}+\left[3 \gamma ^2+2\omega _1^2 \left(1+\varepsilon^2\right)\right]\lambda^{4}+\gamma  \left[\gamma ^2+4\omega _1^2 \left(1+\varepsilon^2\right)\right] \lambda^{3}+
\\\label{caratt_3}
\nonumber &\left[3 \gamma ^2\omega _1^2 \left(1+\varepsilon^2\right)+\omega _1^4 \left(1-\varepsilon^2\right)^2\right]\lambda^{2}+ \gamma  \left[\gamma ^2 \omega _1^2 \left(1+\varepsilon^2\right)+\omega _1^4 \left(1-\varepsilon^2\right)^2\right]\lambda^{1}
-\gamma ^2  \omega _1^4\varepsilon^2 =0.
\end{align}


For the approximated solutions, we get:

\begin{eqnarray} \label{z1} 
&&\lambda_1=\varepsilon^2\frac{\gamma  \omega _1^2}{\gamma ^2+\omega _1^2},
\\  \label{z2} 
&&\lambda_2=-\gamma-\varepsilon^2\frac{\gamma  \omega _1^2}{\gamma ^2+\omega _1^2},
\\\label{z3}
&&\lambda_3 =-\gamma +i \omega _1+\varepsilon^2\left[\frac{\omega _1^2}{\gamma }+\frac{1}{2}\frac{i \gamma  \omega _1}{  \gamma - i \omega _1}\right],
\\\label{z4}
&&\lambda_4=-\gamma -i \omega _1+\varepsilon^2\left[\frac{\omega _1^2}{\gamma }-\frac{1}{2}\frac{i \gamma  \omega _1}{  \gamma +  i \omega _1}\right],
\\\label{z5}
&&\lambda_5 =i\omega_1+\varepsilon^2\left[\frac{1}{2}\frac{i \gamma  \omega _1}{ \gamma + i \omega _1}-\frac{\omega _1^2}{\gamma }\right]
\\\label{z6}
&&\lambda_6 =-i\omega_1+\varepsilon^2\left[-\frac{1}{2}\frac{i \gamma  \omega _1}{ \gamma - i \omega _1}-\frac{\omega _1^2}{\gamma }\right]
\end{eqnarray}




\begin{thebibliography}{}

\bibitem{Fermi} E. Fermi, Phys. Rev. {\bf 75}, 1169 (1949); E. Fermi, The Astrophys. J. {\bf 119}, 1-6 (1954).
\bibitem{arbell}A.R. Bell, Monthly Notices of the Royal Astronomical Society, {\bf 182},  2,  147–156, (1978).
\bibitem{davids}R. C. Davidson, Methods in Nonlinear Plasma Theory (1972)
\bibitem{trub} B A Trubnikov 1997 Phys.-Usp. {\bf40}, 325 (1997).
\bibitem{drury} L. O'C. Drury, Rep. on Progr. in Phys. {\bf 46} (8), 973 (1983).
\bibitem{vulpe} F. Bouchet, F. Cecconi, A. Vulpiani, Phys. Rev. Lett. {\bf 92}, 040601 (2004).
\bibitem{moine} M. Lemoine, Phys. Rev. D {\bf 99}, 083006 (2019); M. Lemoine Phys. Rev. Lett. {\bf  129}, 215101 (2022).


\bibitem{an} Q. An {\it et al.},  Sci. Adv. v. 5, Issue 9 (2019).
\bibitem{tang} H Tang {\it et al}, Sci. Adv.  v 11, Issue 7  (2025).
\bibitem{comm} S. Raptis {\it et al},  Nature Communications v. 16, 488 (2025)
\bibitem{interm} A. Shukurov {\it et al}, The Astrophys. J.,\textbf{839}, L16 (2017).
\bibitem{interm2} D. Tharakkal, A. P. Snodin, G. R. Sarson, A. Shukurov  Phys. Rev. E 107, 065206  (2023)


\bibitem{marc}Neuer, M.; Spatschek, K.H. Diffusion of test particles in stochastic magnetic fields for small Kubo numbers.  \textit{Phys. Rev. E} \textbf{2006}, \emph{73}, 026404.

\bibitem{a}Shalchi, A. Perpendicular Transport of Energetic Particles in Magnetic Turbulence. \textit{Space Sci. Rev.} \textbf{2020}, \emph{216},  23.
\bibitem{shun}S. Ogawa  {\it et al.}, Phys. Plasmas \emph{23}, 072506 (2016).

\bibitem{met}R. Metzler, J. Klafter, Phys. Rep.   {\bf 339}, 1-77 (2000).
\bibitem{aq3}G. Aquino, K. J. Chand\'ia, M. Bologna, \textit{Entropy} {\bf 23}, 781 (2021).


\bibitem{mocte} R E Moctezuma et al., Phys. Scr. {\bf 100}, 055231 (2025)
\bibitem{pnas}R. Shinde, J. U. Sommer, H.Löwen, A. Sharma, Strongly enhanced dynamics of a charged Rouse dimer by an external magnetic field, PNAS Nexus, {\bf 1}, Issue 3, pgac119 (2022).
\bibitem{jap} N.Nakayama, H. Kawamoto, S. Yamada, Resonance Frequency and Stiffness of Magnetic Bead Chain in Magnetic Field, Journal of Imaging Science and Technology,   pp 408 - 417,  (2003).
\bibitem{supermag} E. Rapoport and G. S. D. Beach, J. Appl. Phys. {\bf 111}, 07B310 (2012)
\bibitem{kavorkia} J. Kevorkian , J. D. Cole, Perturbation Methods in Applied Mathematics, pag. 316, 1996 Springer-Verlag NewYork, Inc. 
\bibitem{bolo} M. Bologna, Exact Approach to Uniform Time-Varying Magnetic Field, Mathematical Problems in Engineering, (2018) Article ID 9521975.


\bibitem{land8}Landau, L.D.; Lifshitz, E.M.; Pitaevskii, L.P.
\emph{Electrodynamics of Continuous Media}, 2nd ed.;    Elsevier Butterworth-Heinemann: Oxford, UK,  1984.

\bibitem{jak}Jackson, J.D. \emph{Classical Electrodynamics}, 3rd ed.; {John Wiley\&Sons:  Hoboken, NJ, USA, 
} 1999.

\bibitem{gitt2}Gitterman, M. \emph{The Noisy Oscillator The First Hundred Years, From Einstein Until Now};  {World Scientific Publishing: Singapore, 
}2005.
\bibitem{gitt} S. Burov, M. Gitterman,  Phys. Rev. E  \emph{94}, 052144 (2016).
\bibitem{log}V. E. Shapiro, V.M  Loginov,   Phys. A  {\bf 91}, 563-574 (1978).
\bibitem{bel} G. Bel, E. Barkai, E. Phys. Rev. Lett.  {\bf 94}, 240602 (2005).
\bibitem{aq2}G. Aquino, L. Palatella, P.  Grigolini,  Phys. Rev. Lett.  \emph{93}, 050601 (2004).
\bibitem{prl} G. Aquino, M. Bologna, P. Grigolini, B. West, Phys. Rev. Lett. {\bf 105}, 040601 (2010).
%
\bibitem{pnasB} S. Burov, E. Metzler, E. Barkai, Proceedings of the National Academy of Sciences {\bf 107} (30), 13228-13233 (2010).
\bibitem{barka} S. Burov, and E. Barkai,  Physical Review E, {\bf 78}, 031112  (2008).
\bibitem{west} B.J. West, M Bologna, P. Grigolini, Physics of Fractal Operators, Springer, (2003).
\bibitem{superp} A. L. Pankratov  et al. , Nature Communications  {\bf 16}, 3457 (2025).
\bibitem{glass1}B. Riechers, A. Das, E. Dufresne et al. Intermittent cluster dynamics and temporal fractional diffusion in a bulk metallic glass. Nat Commun {\bf 15}, 6595 (2024).
\bibitem{glass2}G. Aquino, A. Allahverdyan, T. M. Nieuwenhuizen, Phys. Rev. Lett. {\bf 101}, 015901(2008).
\bibitem{glass3} T.Kleiner, R.Hilfer, Fractional glassy relaxation and convolution modules of distributions. Anal.Math.Phys. {\bf 11}, 130 (2021). 

\bibitem{Manneville}P. Manneville,  J. Physique {\bf 41}, 1235–1243 (1980).
\bibitem{aq4}G. Aquino, N Scafetta and P. Grigolini  \textit{Chaos Solitons \& Fractals } \textbf{2001}, \emph{12 (11)}, 2023-2038.




\bibitem{zumofenklafter}G. Zumofen and  J. Klafter, Phys. Rev. E {\bf 47}, 851(1993).
\end{thebibliography}
\end{document}